\documentclass[twocolumn,twoside]{IEEEtran}

\usepackage{graphicx,epic,eepic,epsfig,amsmath,latexsym,amssymb,verbatim,color,revsymb}
\usepackage{theorem}
\usepackage{cite}

\usepackage[usenames,dvipsnames]{pstricks}

\newtheorem{definition}{Definition}
\newtheorem{proposition}[definition]{Proposition}
\newtheorem{lemma}[definition]{Lemma}

\newtheorem{theorem}[definition]{Theorem}
\newtheorem{corollary}[definition]{Corollary}

\def\squareforqed{\hbox{\rlap{$\sqcap$}$\sqcup$}}
\def\qed{\ifmmode\squareforqed\else{\unskip\nobreak\hfil
\penalty50\hskip1em\null\nobreak\hfil\squareforqed
\parfillskip=0pt\finalhyphendemerits=0\endgraf}\fi}
\def\endenv{\ifmmode\;\else{\unskip\nobreak\hfil
\penalty50\hskip1em\null\nobreak\hfil\;
\parfillskip=0pt\finalhyphendemerits=0\endgraf}\fi}
\newcommand{\altqed}{\hfill{\small $\blacksquare$}}
\newenvironment{remark}{\noindent \textbf{{Remark~}}}{\altqed}
\newenvironment{example}{\noindent \textbf{{Example~}}}{\altqed}

\mathchardef\ordinarycolon\mathcode`\:
\mathcode`\:=\string"8000
\def\vcentcolon{\mathrel{\mathop\ordinarycolon}}
\begingroup \catcode`\:=\active
  \lowercase{\endgroup
  \let :\vcentcolon
  }

\newcommand{\nc}{\newcommand}
\nc{\rnc}{\renewcommand}
\nc{\beq}{\begin{equation}}
\nc{\eeq}{{\end{equation}}}
\nc{\beqa}{\begin{eqnarray}}
\nc{\eeqa}{\end{eqnarray}}
\nc{\beas}{\begin{eqnarray*}}
\nc{\eeas}{\end{eqnarray*}}
\nc{\lbar}[1]{\overline{#1}}
\nc{\bra}[1]{\langle#1|}
\nc{\ket}[1]{|#1\rangle}
\nc{\ketbra}[2]{|#1\rangle\!\langle#2|}
\nc{\braket}[2]{\langle#1|#2\rangle}
\nc{\proj}[1]{| #1\rangle\!\langle #1 |}
\nc{\avg}[1]{\langle#1\rangle}
\nc{\Rank}{\operatorname{Rank}}
\nc{\smfrac}[2]{\mbox{$\frac{#1}{#2}$}}
\nc{\tr}{\operatorname{Tr}}
\nc{\ox}{\otimes}
\nc{\dg}{\dagger}
\nc{\dn}{\downarrow}
\nc{\cA}{{\cal A}}
\nc{\cB}{{\cal B}}
\nc{\cC}{{\cal C}}
\nc{\cD}{{\cal D}}
\nc{\cE}{{\cal E}}
\nc{\cF}{{\cal F}}
\nc{\cG}{{\cal G}}
\nc{\cH}{{\cal H}}
\nc{\cI}{{\cal I}}
\nc{\cJ}{{\cal J}}
\nc{\cK}{{\cal K}}
\nc{\cL}{{\cal L}}
\nc{\cM}{{\cal M}}
\nc{\cN}{{\cal N}}
\nc{\cO}{{\cal O}}
\nc{\cP}{{\cal P}}
\nc{\cR}{{\cal R}}
\nc{\cS}{{\cal S}}
\nc{\cT}{{\cal T}}
\nc{\cX}{{\cal X}}
\nc{\cZ}{{\cal Z}}
\nc{\csupp}{{\operatorname{csupp}}}
\nc{\qsupp}{{\operatorname{qsupp}}}
\nc{\var}{{\operatorname{var}}}
\nc{\rar}{\rightarrow}
\nc{\lrar}{\longrightarrow}
\nc{\polylog}{{\operatorname{polylog}}}
\nc{\1}{{\openone}}
\nc{\wt}{{\operatorname{wt}}}
\nc{\av}[1]{{\left\langle {#1} \right\rangle}}

\nc{\RR}{{{\mathbb R}}}
\nc{\CC}{{{\mathbb C}}}
\nc{\FF}{{{\mathbb F}}}
\nc{\NN}{{{\mathbb N}}}
\nc{\ZZ}{{{\mathbb Z}}}
\nc{\PP}{{{\mathbb P}}}
\nc{\QQ}{{{\mathbb Q}}}
\nc{\UU}{{{\mathbb U}}}
\nc{\EE}{{{\mathbb E}}}
\nc{\id}{{\operatorname{id}}}

\nc{\CHSH}{{\operatorname{CHSH}}}

\nc{\be}{\begin{equation}}
\nc{\ee}{{\end{equation}}}
\nc{\bea}{\begin{eqnarray}}
\nc{\eea}{\end{eqnarray}}
\nc{\<}{\langle}
\rnc{\>}{\rangle}
\nc{\Hom}[2]{\mbox{Hom}(\CC^{#1},\CC^{#2})}
\nc{\rU}{\mbox{U}}

\nc{\ob}[1]{#1}

\nc{\SEP}{{\text{SEP}}}
\nc{\sep}{{\text{sep}}}
\nc{\LOCC}{{\text{LOCC}}}
\nc{\PPT}{{\text{PPT}}}
\nc{\EXT}{{\text{EXT}}}
\nc{\Sym}{{\operatorname{Sym}}}
\nc{\SWAP}{{\operatorname{SWAP}}}

\newcommand{\bes}{\begin{equation*}}
\newcommand{\ees}{\end{equation*}}

\newcommand{\threshold}{{\frac{1}{\sqrt{2}}}}
\newcommand{\semistrong}{{pretty strong}}
\newcommand{\Semistrong}{{Pretty strong}}

\def\tr{{\rm Tr}\,}

\begin{document}

\title{``\Semistrong'' converse for the quantum capacity\protect\\ of degradable channels}

\author{Ciara Morgan and Andreas Winter%
\thanks{Date: 28 October 2013.}%
\thanks{CM is with the Institut f\"ur Theoretische Physik, Leibniz Universit\"at Hannover, Appelstra\ss{}e 2, D-30167 Hannover, Germany. Email: {\tt ciara.morgan@itp.uni-hannover.de}}%
\thanks{AW is with ICREA and F\'{\i}sica Te\`{o}rica: Informaci\'{o} i Fenomens Qu\`{a}ntics, Universitat Aut\`{o}noma de Barcelona, ES-08193 Bellaterra (Barcelona), Spain. He is currently on leave from the
Department of Mathematics, University of Bristol, Bristol BS8 1TW, U.K. Email: {\tt andreas.winter@uab.cat}}%
\thanks{This work was started when both authors were with CQT, National University of Singapore, 3 Science Drive 2, Singapore 117543.}%
\thanks{CM is funded from the EU grant QFTCMPS and the cluster of excellence EXC 201 Quantum Engineering and Space-Time Research. %
AW acknowledges financial support by the European Commission (STREPs ``QCS'' and ``RAQUEL'', and Integrated Project ``QESSENCE''), the ERC (Advanced Grant ``IRQUAT''), a Royal Society Wolfson Merit Award and a Philip Leverhulme Prize. Furthermore, by the Spanish MINECO, project FIS2008-01236, with the support of FEDER funds. The Centre for Quantum Technologies is funded by the Singapore Ministry of Education and the National Research Foundation as part of the Research Centres of Excellence programme.}}

\maketitle

\begin{abstract}
We exhibit a possible road towards a strong converse for the quantum 
capacity of degradable channels.
In particular, we show that all degradable channels obey what we call a 
``\semistrong{}'' converse: When the code rate increases above the 
quantum capacity, the fidelity makes a discontinuous jump from $1$
to at most $\frac{1}{\sqrt2}$, asymptotically. 
A similar result can be shown for the private (classical) capacity.
\par
Furthermore, we can show that if the strong converse holds for symmetric 
channels (which have quantum capacity zero),
then degradable channels 
obey the strong converse: The above-mentioned asymptotic jump of 
the fidelity at the quantum capacity is then from $1$ down to $0$.
\end{abstract}


\maketitle


\section{Introduction}
Communication via noisy channels is one of the information processing tasks by which,
following the fundamental work of Shannon~\cite{Shannon48}, we have learned
to quantify information and noise. One of the most important models considered from
these early days of information theory is that of a discrete memoryless channel,
for which Shannon gave his famous single-letter formula for the capacity
(i.e.,~the maximum communication rate achievable by asymptotically error-free 
block coding). 

The analogous model in quantum Shannon theory is the memoryless quantum
channel $\cN^{\ox n}$ (for asymptotically large integer $n$), 
given by a completely positive and trace preserving (cptp) map 
$\cN:\cL(A') \rightarrow \cL(B)$, with Hilbert spaces $A'$ and $B$ that
we assume to be finite dimensional throughout this paper.

The quantum capacity $Q(\mathcal{N})$ of $\mathcal{N}$ is informally 
defined as the maximum rate at which quantum information can be transmitted
asymptotically faithfully over that channel, when using it $n \rightarrow \infty$
times.

As for all channel capacity theorems, the quantum capacity theorem consists of a
direct part and a converse. The direct part states that for rates below a certain
threshold there exist codes with decoding error (quantified as a certain distance
from noiseless transmission) tending to $0$ in the number of channel uses.
The converse states that if the rate lies above this threshold then the error does
not go to $0$ for any sequence of codes. To be precise, this is known as a
\emph{weak converse} and the threshold rate sometimes called \emph{weak capacity}.
A \emph{strong converse} is the statement that for rates above the capacity
the error converges to its maximum $1$ as $n \to \infty$. 

While the strong converse is not known for the quantum capacity of 
any non-trivial channel (however, see the examples and remarks below in 
Section~\ref{sec:converses}),
strong converse theorems have been shown to hold for other types of information 
sent over memoryless quantum channels, including classical information encoded into 
product states \cite{ON99,Winter99} and for general input states (i.e. allowing the 
possibility of entangled input signal states) over certain classes of quantum 
channels, by \cite{KW09}. The strong converse holds also for
entanglement-assisted classical communication over memoryless quantum channels,
by the Quantum Reverse Shannon Theorem~\cite{QRST,simple-QRST}; the optimal
rate is the entanglement-assisted (classical) capacity, denoted $C_E$~\cite{BSST}.
Strong converses do not hold by default; certain quantum channels with memory have 
a weak capacity but fail the strong converse \cite{DHB11,DM11}.

The paper is structured as follows: In Section~\ref{sec:Q}
we recall the definition of codes, error criteria and the quantum capacity.
Then, in Section~\ref{sec:converses} we discuss the weak converse for the 
quantum capacity and the possibility of strong converses. 
In Section~\ref{sec:degradable}, we review the concept of degradable channels
and the analysis of Devetak and Shor~\cite{DS05} of their quantum
capacity. We will present the argument in a form that will aid in the 
subsequent finer analysis, proving a structural lemma on degradable
channels along the way.
Then in Section~\ref{sec:main}, we state and prove our first main result
(Theorem~\ref{thm:Q-semi-strong}) strongly bounding the rate of channels with
sufficiently small error. All necessary auxiliary results are stated 
in this section, however the proofs are relegated to the appendix.
Subsequently, we prove an analogous rate bound for the private classical
capacity (Theorem~\ref{thm:P-semi-strong} in Section~\ref{sec:P-semi-strong}), 
and then show that a strong converse for all symmetric channels implies the strong 
converse for all degradable channels (Theorem~\ref{thm:Q-strong:from-symmetric-to-degradable} 
in Section~\ref{sec:from-symmetric-to-degradable}). 
In Section~\ref{sec:SDP}
we discuss a semidefinite programming approach to deal with the symmetric channels.
We conclude in Section~\ref{sec:conclusion} with a brief discussion of
what was achieved and highlight open problems.

\section{Quantum channel capacity}
\label{sec:Q}
For a given channel $\cN:\cL(A')\rightarrow\cL(B)$, we consider encoding
and decoding of quantum information, given by completely positive and trace preserving (cptp) maps
\begin{align*}
  \cE &: \cL(C) \rightarrow \cL(A'), \\
  \cD &: \cL(B) \rightarrow \cL(C),
\end{align*}
which together form a \emph{quantum code}. The idea is that the information 
to be sent is subjected to the overall effective
channel $\cD\circ\cN\circ\cE:\cL(C) \rightarrow \cL(C)$. 
For a Hilbert space $\cH$, we denote  by
\begin{align*}
  \cS(\cH)        &= \{ \rho \geq 0 \text{ s.t. } \tr\rho = 1\},  \\
  \cS_{\leq}(\cH) &= \{ \rho \geq 0 \text{ s.t. } \tr\rho \leq 1\},
\end{align*}
the set of states and sub-normalized densities, respectively.

There are many ways of defining mathematically the notion that the output is a
good approximation of the input, and we refer the reader to the comprehensive
treatment of Kretschmann and Werner~\cite{tema-con-variazioni} for a discussion
of all the concomitant ways of defining the capacity and the proof that
asymptotically and for vanishing error they are the same.
In the present paper we will measure the degree of approximation between 
states by the fidelity, given as
\[
  F(\rho,\sigma) := \left\| \sqrt{\rho}\sqrt{\sigma} \right\|_1
                 =  \max |\bra{\varphi}\psi\rangle|,
\]
where the maximization is over all purifications $\ket{\varphi}$, $\ket{\psi}$ 
of $\rho$ and $\sigma$, respectively~\cite{Uhlmann:fid,Jozsa:fid}.
This definition extends to subnormalized density operators 
$\rho,\sigma \in \cS_{\leq}(\cH)$ by letting
\[\begin{split}
  F(\rho,\sigma) &:= F\left( \rho \oplus (1-\tr\rho), \sigma \oplus (1-\tr\sigma) \right) \\
                 &= \left\| \sqrt{\rho}\sqrt{\sigma} \right\|_1 + \sqrt{(1-\tr\rho)(1-\tr\sigma)}.
\end{split}\]
It can be shown that both
\begin{align*}
  P(\rho,\sigma) &:= \sqrt{1-F(\rho,\sigma)^2}, \text{ and} \\
  A(\rho,\sigma) &:= \arccos F(\rho,\sigma)
                   = \arcsin P(\rho,\sigma),
\end{align*}
called the \emph{purified distance} and the \emph{geodesic distance}, respectively,
are metrics on $\cS_{\leq}(\cH)$, cf.~\cite{TomamichelThesis}. They are obviously equivalent,
and can be shown to be equivalent to the trace norm distance~\cite{Fuchs-vandeGraaf}:
\begin{equation}
  \label{eq:P-vs-D}
  \frac12 \| \rho-\sigma \|_1 \leq P(\rho,\sigma) \leq \sqrt{\| \rho-\sigma \|_1}.
\end{equation}

In the subsequent definitions, we will consistently use the purified distance.
For instance, the error of a code $(\cE,\cD)$ for $\cN$ is defined as
\[
  P(\id,\cD\circ\cN\circ\cE) := \sup_{C'} \sup_{\rho \in \cS(CC')} P(\rho,(\id\ox\cD\circ\cN\circ\cE)\rho).
\]
The maximum dimension $|C|$ of $C$ such that there exists a quantum 
code for $\cN^{\ox n}$ with error $\epsilon$, is denoted 
$N(n,\epsilon)$, or more precisely $N(n,\epsilon|\cN)$ if we want to 
refer explicitly to the channel.

If we have a code with error $\leq \epsilon$, this means that we can use it with
the maximally entangled state $\ket{\Phi}^{CC'}$ at the input, to get an
output state
\[
  \sigma^{CC'} = (\id \ox \cD\circ\cN\circ\cE)\Phi
               = (\id \ox \cD\circ\cN)(\id \ox\cE)\Phi,
\]
which is $\epsilon$-close to being maximally entangled: $P(\Phi,\sigma) \leq \epsilon$. This
motivates the definition of an \emph{entanglement-generating code with error $\epsilon$},
which consists of a state $\rho^{A'C'}$ and a decoding cptp map 
$\cD:\cL(B) \rightarrow \cL(C)$, such that
\[
  P(\Phi^{CC'},(\id \ox \cD\circ\cN)\rho^{A'C'}) \leq \epsilon.
\]
The maximum dimension $|C|$ of $C$ such that there exists an entanglement-generating
code for $\cN^{\ox n}$ with error $\epsilon$, is denoted $N_E(n,\epsilon)$, or more explicitly,
$N_E(n,\epsilon|\cN)$. Clearly, $N(n,\epsilon) \leq N_E(n,\epsilon)$. 

\medskip
\begin{remark}
Since the purified distance 
$P(\Phi,(\id \ox \cD\circ\cN)\rho) = \sqrt{1-\tr\bigl((\id \ox \cD\circ\cN)\rho\bigr)\Phi}$
is concave in $\rho$, we may always assume that the state $\rho$ on $A'C'$ in an
entanglement-generating code is pure, as in each convex decompositions of $\rho$ there 
is at least one state with an error no larger than that of $\rho$.
\end{remark}

\medskip
The quantum capacity is now defined as
\[
  Q(\cN) = \inf_{\epsilon > 0} \liminf_{n\rightarrow\infty} \frac{1}{n} \log N(n,\epsilon).
\]
One obtains the same capacity when using $\limsup$ and $N_E$, see~\cite{tema-con-variazioni} 
for a proof of this and the equivalence of other variations of the definition.
On notation: In this paper, $\log$ is always the binary logarithm, and 
$\exp$ its inverse, the exponential function to base $2$. The natural logarithm
is denoted $\ln x$, the natural exponential function $e^x$.

\medskip
A Shannon-style formula for the quantum capacity was
first stated by Lloyd \cite{Lloyd} and proved rigorously by Shor \cite{Shor} and
Devetak \cite{Devetak}. More precisely, in these papers they prove the direct (achievability)
part which together with the earlier result of  Schumacher and Nielsen \cite{SN96,SN96a}, who showed the same quantity to be an upper bound (i.e.,~weak converse), leads to a formula for the quantum capacity. We expand upon this weak converse in the following section.

The formula for the quantum capacity is given in terms of the \emph{coherent information}
\[
  I(A \rangle B)_{\rho} = -S(A|B)_\rho = S(\rho^B) - S(\rho^{AB}),
\]
where $S(\rho) = - \tr\rho\log\rho$ is the von Neumann entropy, of a state
$\rho^{AB} = (\id\ox\cN)\phi^{AA'}$ with a ``test state'' $\phi$ on $AA'$.
Namely,
\[
  Q(\mathcal{N}) = \lim_{n \rightarrow \infty} \frac{1}{n} Q^{(1)}(\cN^{\ox n}),
\]
with the single-letter expression
\[
  Q^{(1)}(\cN) = \max_{\phi\in\cS(AA')} \{ I(A \rangle B)_\rho : \rho = (\id\ox\cN)\phi \}.
\]

\medskip
\begin{remark}
The quantum capacity is known to be non-additive \cite{SY08}. So is the 
single-letter quantity $Q^{(1)}(\cN)$~\cite{ShorSmolin,DiVincenzoShorSmolin}, 
meaning that the regularization above is necessary, at least as long as 
we base our capacity formula on the coherent information. 
It is not known whether there is a single-letter 
formula for $Q(\cN)$, or even an efficient approximation scheme~\cite{Shor:caps}.
As a matter of fact, we do not even know how to characterize the
quantum capacity of the qubit depolarizing channel as a function of
the noise, the currently best upper bounds being those by
Ouyang~\cite{Ouyang}, the best lower bounds are due to Fern and
Whaley~\cite{FernWhaley}.
\end{remark}

\section{Weak and strong converse}
\label{sec:converses} 
The fact that the coherent information gives an upper bound on the quantum
capacity of general channels has been known since Schumacher and Nielsen~\cite{SN96}.
They showed that for any entanglement generating code with code space $C$,
for a channel $\cN:\cL(A')\rightarrow \cL(B)$ with error $\epsilon$, using 
strong subadditivity together with Eq.~(\ref{eq:P-vs-D}) and the Fannes inequality,
there exists an input test state $\phi^{AA'}$ such that with 
$\rho^{AB} = (\id\ox\cN)\phi$,
\[
  (1-2\epsilon)\log |C| \leq I(A\rangle B)_\rho + 1.
\]
Applying this to a maximal code for $\cN^{\ox n}$ yields, for $\epsilon < \frac12$,
\begin{equation}
  \label{eq:weak}
  \frac{1}{n} \log N_E(n,\epsilon) 
     \leq \frac{1}{1-2\epsilon} \frac{1}{n}Q^{(1)}(\cN^{\ox n}) + \frac{1}{(1-2\epsilon)n},
\end{equation}
hence the result that for $n\rightarrow\infty$ and $\epsilon \rightarrow 0$,
the optimal rate cannot exceed $\lim_n \frac{1}{n}Q^{(1)}(\cN^{\ox n})$, which
we know is also asymptotically achievable, thanks to Lloyd-Shor-Devetak.

However, for any non-zero $\epsilon > 0$, the upper bound in Eq.~(\ref{eq:weak})
is a constant factor away from the capacity, which is the hallmark of a 
weak converse; it leaves room for a trade-off between 
communication rate and error, asymptotically.

If the quantum capacity $Q(\cN)$ is zero, Eq.~(\ref{eq:weak}) says something
a bit stronger, namely that $N_E(n,\epsilon) \leq O(1)$, at least when 
$\epsilon < \frac12$. In this article we call such a statement \emph{\semistrong{}
converse}, i.e.~a proof amounting to
\[
  \limsup_{n\rightarrow\infty} \frac{1}{n}\log N_E(n,\epsilon) \leq Q(\cN),
\]
at least for error $\epsilon$ below some threshold $\epsilon_0$. 
By the preceding argument, channels with vanishing capacity obey 
a \semistrong{} converse.
A strong converse would require the above for all $\epsilon < 1$;
cf.~\cite[Sec.~2.7]{tema-con-variazioni}.

\medskip
Here are two simple examples of channels for which the strong converse holds.

\medskip
\begin{example}
\emph{(PPT entanglement binding channels).} 
If $\cN$ is such that all $\rho = (\id\ox\cN)\phi$ have positive
partial transpose (PPT), then any entanglement generating code for a 
maximally entangled state of Schmidt rank $d$, denoted $\Phi_d$, using any
number $n$ of channel uses and even arbitrary classical communication
on the side, can only generate a PPT state between the communicating
parties. Twirling by the symmetries
$U\ox\overline{U}$ of the maximally entangled state does not
change the fidelity between the resulting state and the maximally
entangled state. But the resulting isotropic state
\[
  \rho = p\Phi_d + (1-p)\frac{1}{d^2-1}(\1-\Phi_d)
\]
is still PPT, and it is well-known that this can only hold for 
$p\leq \frac{1}{d}$~\cite{Rains:SDP}.
I.e., the error is at least $\sqrt{1-\frac{1}{d}}$, which in the
setting of $n$ channel uses ($\cN^{\ox n}$) goes to $1$ exponentially 
fast for positive rates (meaning $d=2^{nR}$ with $R>0$).
\end{example}

\medskip
\begin{example}
\emph{(Ideal channel).} Consider the identity $\id_2:\cL(\CC^2)\rightarrow\cL(\CC^2)$
on a qubit and an entanglement-generating code for $n$ uses of it,
$\id_2^{\ox n}$ for a maximally entangled state of rank $d$. It is evident
that the state shared between sender and receiver after the transmission is
of Schmidt rank $\leq 2^n$, and so is any state obtained by the receiver's
decoding. Hence the fidelity of the code is upper bounded by
\[
  \max \big\{ |\bra{\Phi_d} \psi\rangle|\, : \,\text{Schmidt rank of } \ket{\psi} 
                                               \text{ at most } 2^n \bigr\}
                                                            = \sqrt{\frac{2^n}{d}}.
\]
Consequently, as soon as the rate is above the capacity $Q(\id_2) = 1$, 
i.e.~$d=2^{nR}$ for $R>1$, the error goes to $1$ exponentially fast.
\end{example}

\medskip
\begin{remark}
At this juncture we should point out that for any channel $\cN$, 
and for sufficiently large rates $R>R_0$, one can prove that
the error is going to $1$, even exponentially fast. (However, 
we do not call this a strong converse
for the channel, unless $R_0$ equals the quantum capacity $Q$.)

All known proofs of this statement are based on simulation of the
channel by a limited rate $R_0$ of the ideal channel, with unrestricted
encodings and decodings, and possibly including some other extra
free resource that does not change the capacity of the ideal channel.
This is because the local parts of the simulation can be absorbed 
into a potential transmission code for the channel, and the the ideal
channel example above applies.

With free entanglement the rate is $Q_E(\cN) = \frac12 C_E(\cN)$, 
the entanglement-assisted quantum capacity, by the 
Quantum Reverse Shannon Theorem~\cite{BSST,QRST,simple-QRST}.
With free classical communication the rate is $E_C(\cN)$, the
\emph{entanglement cost} of the channel~\cite{channel-E-cost}. Both rates
are upper bounds on $Q(\cN)$, the latter even on the two-way
classical-communication-assisted quantum capacity $Q_2(\cN)$, 
they are known to be incomparable (meaning that there are cases where
either can be much better than the other) and generally not tight.
For instance, consider any PPT entanglement-binding channel,
for which the first example above shows that the strong converse
holds, with quantum capacity $Q=0$. However, both of the mentioned
simulations of the channel guarantee error convergence to $1$ only
at rates $Q_E, E_C > 0$. Indeed, $Q_E=0$ if and only if the channel
were constant, and $E_C=0$ if and only if the channel were
entanglement-breaking~\cite{channel-E-cost,Yang-et-al}.
\end{remark}

\section{Degradable and anti-degradable channels}
\label{sec:degradable}
By the Stinespring dilation theorem, any channel can be defined by an isometric embedding
$U : A' \longrightarrow B \otimes E$ followed by a partial trace over the environment system $E$,
such that $\mathcal{N}(\rho) = \tr_E U \rho U^\dagger$. Tracing over $B$ rather than $E$ we
obtain the corresponding complementary channel, $\mathcal{N}^c(\rho) := \tr_B U \rho U^\dagger$.

As we are interested in the channel's behaviour, we will without loss
of generality assume from now on that $E$ is chosen to be of minimal dimension
(which makes $U$ unique up to isometries on $E$). Furthermore, since $\cN$ is
the complementary channel of $\cN^c$, we may equally reduce the dimension
of $B$ if needed; this can equivalently be described as finding the subspace 
$\widehat{B} \subset B$ that contains all supports of all $\cN(\rho)$ for 
states $\rho$ on $A'$, which is in fact the supporting subspace of $\cN(\1)$,
and viewing $\cN$ as a mapping into $\cL(\widehat{B})$. 

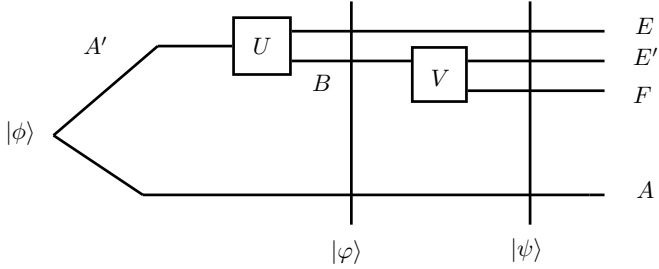
\begin{figure}[ht]
\begin{center}
\scalebox{.9}{
\begin{pspicture}(0,-1.9889063)(10.002812,1.9689063)
\usefont{T1}{ptm}{m}{n}
\rput(3.9614062,1.2989062){$U$}
\usefont{T1}{ptm}{m}{n}
\rput(9.612344,1.5789062){$E$}
\usefont{T1}{ptm}{m}{n}
\rput(9.632344,1.1389062){$E'$}
\usefont{T1}{ptm}{m}{n}
\rput(9.582344,0.5589062){$F$}
\usefont{T1}{ptm}{m}{n}
\rput(1.5023438,1.2989062){$A'$}
\usefont{T1}{ptm}{m}{n}
\rput(9.622344,-0.84109384){$A$} 
\usefont{T1}{ptm}{m}{n}
\rput(4.842344,0.7389062){$B$}
\usefont{T1}{ptm}{m}{n}
\rput(5.212344,-1.7610937){$\ket{\varphi}$}
\usefont{T1}{ptm}{m}{n}
\rput(7.872344,-1.7210938){$\ket{\psi}$}
\usefont{T1}{ptm}{m}{n}
\rput(0.39234376,-0.0010938){$\ket{\phi}$}
\psline[linewidth=0.04cm](2.2,-0.9110938)(8.8,-0.9110938)
\psline[linewidth=0.04cm](2.42,1.2889062)(3.52,1.2889062)
\psframe[linewidth=0.04,dimen=outer](4.4,1.7289063)(3.52,0.8489063)
\psline[linewidth=0.04cm](5.28,1.9489063)(5.28,-1.3510938)
\psframe[linewidth=0.04,dimen=outer](7.0,1.2889062)(6.16,0.44890624)
\psline[linewidth=0.04cm](7.92,1.9489063)(7.92,-1.3510938)
\psline[linewidth=0.04cm](8.8,-0.9110938)(9.02,-0.9110938)
\psline[linewidth=0.04cm](0.88,-0.03109375)(2.42,1.2889062)
\psline[linewidth=0.04cm](0.88,-0.03109375)(2.2,-0.9110938)
\psline[linewidth=0.04cm](4.4,1.5089062)(9.02,1.5089062)
\psline[linewidth=0.04cm](4.4,1.0689063)(6.16,1.0689063)
\usefont{T1}{ptm}{m}{n}
\rput(6.5914063,0.81890625){$V$}
\psline[linewidth=0.04cm](7.0,1.0689063)(9.02,1.0689063)
\psline[linewidth=0.04cm](7.0,0.62890625)(9.02,0.62890625)
\end{pspicture}}
\caption{Schematic of a degradable quantum channel, with the input state $\phi$ between $A'$
and the reference $A$, the channel output and environment state $\varphi$ and the state
$\psi$ shared between $A$, $F$ and the two copies of the original environment, $E$ and $E'$.}
\label{DegradableChannelFig}
\end{center}
\end{figure}

A channel $\mathcal{N}$ is called \emph{degradable} if it can be degraded to
its complementary channel, i.e.~if there exists a cptp map $\mathcal{M}$ such that
$\mathcal{N}^c = \mathcal{M} \circ \mathcal{N}$.
Introducing the Stinespring dilation of $\mathcal{M}$ by an isometry
$V : B \longrightarrow F \ox E'$, the channel output system $B$ can be mapped to the composite
system $E' \otimes F$ such that the channel taking $A'$ to $E$ is the same as
the channel taking $A'$ to $E'$ (with an isomorphism between $E$ and $E'$ fixed once and
for all). We may also assume $F$ to be minimal.
The above information process is illustrated in Fig.~\ref{DegradableChannelFig}. 

If the complementary channel is degradable, i.e.~if $\cN = \cM \circ \cN^c$ for 
some cptp map, we call $\cN$ \emph{anti-degradable}. A channel that is both
degradable and anti-degradable is called \emph{symmetric}~\cite{SSW}.

\medskip
\begin{example}
Many interesting channels are degradable, for instance the erasure channel
\begin{align*}
  \cE_q:\cL(A) &\longrightarrow \cL(A\oplus\CC\ket{*}), \\
        \rho   &\longmapsto (1-q)\rho \oplus q\proj{*},
\end{align*}
for $0\leq q\leq \frac12$; for $\frac12 \leq q \leq 1$ it is
anti-degradable. 

Isotropically depolarizing channels are in general not degradable,
but for sufficiently large noise, they are known to be 
anti-degradable~\cite{BDSW,SSW,Ouyang}.

A very broad class of degradable channels are so-called Hadamard
channels~\cite{King-et-al}, also known as generalized dephasing channels,
the simplest of which is
\begin{align*}
  \cZ_p:\cL(\CC^2) &\longrightarrow \cL(\CC^2), \\
        \rho       &\longmapsto (1-p)\rho + p Z\rho Z,
\end{align*}
with the Pauli $Z$ matrix. This is a channel for which the
quantum capacity is known: $Q(\cZ_p) = 1-H(p,1-p)$~\cite{DS05,Rains:SDP}.
On the other hand, the simulation arguments discussed in Section~\ref{sec:converses}
do not yield the strong converse. Indeed,
$Q_E(\cZ_p) = 1-\frac12 H(p,1-p)$ and
\[\begin{split}
  E_C(\cZ_p) &\geq E_C\bigl( (1-p)\Phi^+ + p\Phi^- \bigr)            \\
             &=    H\left(\frac12 \pm \sqrt{p(1-p)}\right),
\end{split}\]
the latter by~\cite{Wootters,VDC}; both of these bounds are strictly
larger than $Q(\cZ_p)$ for $p\in(0,1)\setminus\{\frac12\}$.
\end{example}

\medskip
The identity between the channels $\cL(A') \rightarrow \cL(E)$ and 
$\cL(A') \rightarrow \cL(E')$ (defined by conjugating by $VU$ and tracing over 
$E'F$ and $EF$, respectively) is expressed by the equation
\begin{equation}
  \label{eq:environments}
  \psi^{AE} = \psi^{AE'},
\end{equation}
modulo the implicit isomorphism between $E$ and $E'$. This was enough for
Devetak and Shor~\cite{DS05} to prove that for degradable channels the coherent information
is additive; see also \cite[Sec.~A.2]{CRS08}.
The crucial point in their argument is that the coherent information can be
rewritten as a conditional entropy,
\begin{equation}
  \label{eq:coherent-info-conditional-entropy}
  I(A\rangle B)_{\varphi} = S(F|E')_{\psi}.
\end{equation}
Then, based on the observation that the state $\psi^{FE'}$ on the r.h.s. is a linear
function of the input state $\rho^{A'} = \tr_{A} \phi$, and using strong subadditivity, 
one gets subadditivity of the coherent information of a product channel,
hence additivity of $Q^{(1)}$. Below we give an alternative account of the
reasoning leading to Eq.~(\ref{eq:coherent-info-conditional-entropy}), 
which while being more complicated than those cited, has the benefit
of suggesting an extension to min-entropies (Section~\ref{sec:main}).
For the class of degradable channels it is also known that the quantum capacity 
equals the private capacity \cite{Smith08} -- see Section~\ref{sec:P-semi-strong}
below.

Denoting $\SWAP_{EE'}$ the swap unitary between systems $E$ and $E'$, 
i.e.~$\SWAP\ket{u}\ket{v} = \ket{v}\ket{u}$ (always modulo the implicit
identification of $E$ with $E'$), we have the following statement
strengthening Eq.~(\ref{eq:environments}):

\begin{lemma}
\label{lemma:degrading}
Consider a degradable channel $\cN$ with Stinespring dilation $U:A' \hookrightarrow B\ox E$.
Then there exists a degrading map $\cM$ with Stinespring dilation 
$V:B \hookrightarrow F\ox E'$ (not necessarily with minimal dimension $|F|$)
and a unitary $X$ on $F$, which may be chosen as an involution
(i.e.~$X^2=\1$), such that 
\[
  (X_F \ox \SWAP_{EE'})VU = VU.
\]
In particular, for arbitrary state vector 
$\ket{\phi}^{AA'}$ and $\ket{\psi}^{AFEE'} := (\1\ox VU)\ket{\phi}$,
\[
  (\1_A \ox X_F \ox \SWAP_{EE'})\ket{\psi}^{AFEE'} = \ket{\psi}^{AFEE'}.
\]
\end{lemma}
\begin{IEEEproof}
Start with an arbitrary dilation $V_0:B\hookrightarrow F_0 \ox E'$ of an arbitrary map $\cM_0$,
and define the following isometry $W:A \hookrightarrow EE'FG$,
\[
  W := \frac{1}{\sqrt{2}} \bigl( V_0U \ox \ket{0}^G + \SWAP_{EE'}V_0U \ox \ket{1}^G \bigr),
\]
with a qubit system $G$. Let $F = F_0 \ox G$ and $X_F := \1_{F_0} \ox X_G$, where
$X$ is the Pauli $\sigma_x$ unitary on $G$. Evidently,
\[
  W = (\SWAP_{EE'} \ox X_F)W,
\]
and also, since $\cN$ is degradable,
\[
  \tr_{E'F} W\rho W^\dagger = \cN^c(\rho).
\]
Hence, the Stinespring dilations $U$ and $W$ are
equivalent; to be precise, there exists an isometry $V:B \hookrightarrow E'F$
such that $W=VU$, and we get $VU = (\SWAP_{EE'} \ox X_F)VU$.
\end{IEEEproof}

\medskip
The following reasoning uses the chain rule identity $S(AB|C) = S(B|C) + S(A|BC)$
of the conditional von Neumann entropy, but no explicit expansion of any
conditional entropy as a difference of two entropies.
Consider a generic input state $\phi^{AA'}$ to $\cN$ and its associated
$\varphi^{ABE}$ and $\psi^{AFEE'}$. Now, by invariance of the conditional
entropy $S(A|B) = S(AB)-S(B)$ under local unitaries and the duality
identity $S(A|B) = -S(A|C)$ with respect to a pure state on $ABC$,
combined with the above lemma,
\[\begin{split}
  I(A\rangle B)_\varphi &= -S(A|B)_\varphi \\
                        &= -S(A|FE')_\psi \\
                        &= S(F|E') - S(AF|E') \\
                        &= S(F|E') + S(AF|E) \\
                        &= S(F|E') + S(AF|E').
\end{split}\]
This shows that $S(AF|E)=0$, and we obtain Eq.~(\ref{eq:coherent-info-conditional-entropy}).

\section{\Semistrong{} converse}
\label{sec:main}
\begin{theorem}
  \label{thm:Q-semi-strong}
  Let $\cN:\cL(A) \rightarrow \cL(B)$ be a degradable channel with finite
  quantum systems $A$ and $B$. Then, there exists a constant $\mu$ such
  that for error $\epsilon < \threshold$ and every integer $n$,
  \[\begin{split}
    \log N(n,\epsilon) &\leq \log N_E(n,\epsilon)                        \\
                       &\leq n Q^{(1)}(\cN) + \mu\sqrt{n\ln\frac{64 n^{|A|^2}}{\lambda^2}} \\
                       &\phantom{\leq n Q^{(1)}(\cN)}
                        + 3|A|^2 \log n + 5 + 5\log\frac{1}{\lambda},    \\
                       &\leq n Q^{(1)}(\cN) + O\left(\sqrt{n\log n}\right),    
  \end{split}\]
  where $\lambda = \frac14\left(\threshold-\epsilon\right)$.
\end{theorem}

Together with the direct part (achievability proved in~\cite{Lloyd,Devetak,Shor})
we thus get:
\begin{corollary}
For a degradable channel $\cN$, the quantum capacity is given by 
\[\begin{split}
  Q(\cN) &= \lim_{n\rightarrow\infty} \frac{1}{n}\log N(n,\epsilon)  \\
         &= \lim_{n\rightarrow\infty} \frac{1}{n}\log N_E(n,\epsilon),
\end{split}\]
for any $0 < \epsilon < \threshold$. Compared to the original definition this
is simpler as we do not need to vary $\epsilon$, and there is convergence
rather than reference to $\liminf$ or $\limsup$.
\altqed
\end{corollary}

The proof of this theorem will rely on the calculus of min- and max-entropies,
of which we will briefly review the necessary definitions and properties; we refer
the reader to~\cite{TomamichelThesis} for more details.

\begin{definition}[Min- and max-entropy]
For $\rho^{AB} \in \cS_{\leq}(AB)$,
the min-entropy of $A$ conditioned on $B$ is defined as
\[
  H_{\min}(A|B)_{\rho} := \max_{\sigma_B \in \mathcal{S}(B)} 
                          \max \{\lambda \in \mathbb{R} : \rho^{AB} \leq 2^{-\lambda} \1 \otimes \sigma^B \}.
\]
With a purification $\ket{\psi}^{ABC}$ of $\rho$, we define
\[
  H_{\max}(A|B)_{\rho} := - H_{\min}(A|C)_{\psi^{AC}},
\]
with the reduced state $\psi^{AC} = \tr_B\, \psi^{ABC}$.
\end{definition}

\begin{definition}[Smooth min- and max-entropy]
\label{def:smoothed}
Let $\epsilon \geq 0$ and $\rho_{AB} \in \mathcal{S}({AB})$. The 
\emph{$\epsilon$-smooth min-entropy of $A$ conditioned on $B$} is defined as
\[
  H^{\epsilon}_{\min}(A|B)_{\rho} 
     := \max_{{\rho'} \approx_{\epsilon} \rho} H_{\min}(A|B)_{\rho'},
\]
where ${\rho'} \approx_{\epsilon} \rho$ means $P(\rho',\rho) \leq \epsilon$ for
$\rho'\in\cS_{\leq}(AB)$.

Similarly,
\[\begin{split}
  H_{\max}^{\epsilon}(A|B)_{\psi} 
     &:= \min_{{\rho'} \approx_{\epsilon} \rho} H_{\max}(A|B)_{\rho'}\\
     &=  - H_{\min}^{\epsilon}(A|C)_{\psi},
\end{split}\]
with a purification $\psi \in \mathcal{S}({ABC})$ of $\rho$.

All min- and max-entropies, smoothed or not, are invariant under local
unitaries and local isometries.
\end{definition}

\begin{lemma}[Monotonicity]
\label{lemma:mono}
For a state $\rho\in\cS(ABC)$ and any $\epsilon \geq 0$,
\begin{align*}
  H_{\min}^\epsilon(A|BC) &\leq H_{\min}^\epsilon(A|B), \\
  H_{\max}^\epsilon(A|BC) &\leq H_{\max}^\epsilon(A|B).
\end{align*}

Since every cptp map can be written as an isometry followed by a partial trace,
this means that for every $\rho\in\cS(AB)$ and cptp map $\cT:\cL(B)\rightarrow\cL(C)$,
\begin{align*}
  H_{\min}^\epsilon(A|B)_\rho &\leq H_{\min}^\epsilon(A|C)_{(\id\ox\cT)\rho}, \\
  H_{\max}^\epsilon(A|B)_\rho &\leq H_{\max}^\epsilon(A|C)_{(\id\ox\cT)\rho}.
\end{align*}
\altqed
\end{lemma}

The following relations generalize the well-known chain rule identity
$S(AB|C) = S(B|C) + S(A|BC)$ for the von Neumann entropy, albeit for min- and
max-entropies it turns into one of a set of inequalities. 
There are eight versions of it~\cite{VDTR12}, of which we cite only the 
two we are going to use.

\begin{lemma}[Chain rules~\cite{DBWR10,VDTR12}]
\label{lemma:chainRule}
Let $\epsilon,\delta\geq 0$, $\eta >0$. Then, with respect to a
state $\rho \in \mathcal{S}({ABC})$,
\begin{equation}\begin{split}
  \label{eq:chain:max-le-max+max}
  H_{\max}^{\epsilon + 2\delta + \eta}(AB|C)
     &\leq H_{\max}^{\delta}(B|C) + H_{\max}^{\epsilon}(A|BC) \\
     &\phantom{=============}
      + \log\frac{2}{\eta^2},
\end{split}\end{equation}
and
\begin{equation}\begin{split}
  \label{eq:chain:max-ge-min+max}
  H_{\max}^{\epsilon}(AB|C)
     &\geq H_{\min}^{\delta}(B|C) + H_{\max}^{\epsilon+2\delta+2\eta}(A|BC) \\
     &\phantom{=============}
      - 3\log\frac{2}{\eta^2}.
\end{split}\end{equation}
\altqed
\end{lemma}

\begin{lemma}[Proposition 5.5 in \cite{TomamichelThesis}]
\label{lemma:min-max-inequality}
Let $\rho \in \mathcal{S}(AB)$ and $\alpha,\beta \geq 0$ 
such that $\alpha+\beta < \frac{\pi}{2}$. Then,
\begin{equation}
\label{MinMaxIneq}
  H_{\min}^{\sin\alpha}(A|B)_{\rho} 
      \leq H_{\max}^{\sin\beta}(A|B)_{\rho} + \log \frac{1}{\cos^2(\alpha+\beta)}.
\end{equation}
For $\epsilon,\delta \geq 0$, $\epsilon + \delta < 1$ this can be relaxed to the 
simpler form
\begin{equation}
\label{MinMaxIneq-simpler}
  H_{\min}^{\epsilon}(A|B)_{\rho} 
      \leq H_{\max}^{\delta}(A|B)_{\rho} + \log \frac{1}{1-(\epsilon+\delta)^2}.
\end{equation}
\altqed
\end{lemma}

\begin{lemma}[Dupuis~\cite{Fred:personal}]
\label{lemma:max-min-inequality}
Let $\rho \in \mathcal{S}(AB)$ and $0 \leq \epsilon \leq 1$. Then,
\begin{equation}
  \label{MaxMinIneq}
  H_{\max}^{\sqrt{1-\epsilon^4}}(A|B)_\rho \leq H_{\min}^{\epsilon}(A|B)_\rho,
\end{equation}
which can be rewritten and relaxed into the form
\begin{equation}\begin{split}
  \label{MaxMinIneq-simpler}
  H_{\max}^{\delta}(A|B)_\rho &\leq H_{\min}^{\sqrt[4]{1-\delta^2}}(A|B)_\rho \\
                              &\leq H_{\min}^{1-{\frac14}\delta^2}(A|B)_\rho,
\end{split}\end{equation}
for $0 \leq \delta \leq 1$.
\altqed
\end{lemma}

\begin{IEEEproof}[Proof of Theorem~\ref{thm:Q-semi-strong}]
Consider an entanglement generation code for $\log N_E(n,\epsilon)$ ebits
of error $\epsilon$ for the channel $\cN^{\ox n}$. As observed in conjunction 
with the definitions, $N(n,\epsilon) \leq N_E(n,\epsilon)$ and w.l.o.g.~the
input state $\phi^{\widetilde{A}:{A'}^n}$ to the entanglement-generating code
is pure (see Remark in Section \ref{sec:Q}) . Similar to Fig.~\ref{DegradableChannelFig}, write
\begin{align*}
  \ket{\varphi}^{\widetilde{A}B^nE^n}    &:= (\1\ox U^{\ox n})\ket{\phi}, \\
  \ket{\psi}^{\widetilde{A}{E'}^nF^nE^n} &:= \bigl(\1 \ox (V\ox\1)^{\ox n}\bigr)\ket{\varphi}.
\end{align*}

By definition, there exists a decoding cptp map $\cD:\cL(B^n) \rightarrow \cL(\widetilde{A}')$,
such that $\sigma = (\id\ox \cD\circ\cN)\phi$ has purified distance $\leq \epsilon$ from
the maximally entangled state $\Phi_{\widetilde{A}\widetilde{A}'}$. Note
that $|\widetilde{A}| = |\widetilde{A}'| = N_E(n,\epsilon)$. 
Hence, by definition of the max-entropy and using its monotonicity under 
cptp maps (Lemma~\ref{lemma:mono}),
\[\begin{split}
  \log N_E(n,\epsilon)
       &\leq -H^\epsilon_{\max}(\widetilde{A}|\widetilde{A}')_\sigma \\
       &\leq -H^\epsilon_{\max}(\widetilde{A}|B^n)_\varphi \\
       &=    -H^\epsilon_{\max}(\widetilde{A}|{E'}^nF^n)_\psi.
\end{split}\]
The latter, by the duality relation (Definition~\ref{def:smoothed}), 
is equal to $H^\epsilon_{\min}(\widetilde{A}|E^n)$,
which relates the coding performance directly to the decoupling principle 
(cf.~\cite{Dupuis:thesis}).
But we shall not use that route and instead invoke the chain rule 
[Lemma~\ref{lemma:chainRule}, Eq.~(\ref{eq:chain:max-le-max+max})],
with $\eta = \lambda = \frac14\left(\threshold - \epsilon\right)$, to continue
\begin{equation}\begin{split}
  \label{eq:still-general}
  \log N_E(n,\epsilon) &\leq H^\lambda_{\max}(F^n|{E'}^n)  \\
                       &\phantom{==}
                        - H^{\epsilon+3\lambda}_{\max}(\widetilde{A}F^n|{E'}^n) + \log\frac{2}{\lambda^2}.
\end{split}\end{equation}
Let us deal with the second term here first: Using duality, and invoking
Lemma~\ref{lemma:min-max-inequality}, Eq.~(\ref{MinMaxIneq}) 
with $\alpha=\beta=\arcsin(\epsilon+3\lambda) < \frac{\pi}{4}$, 
we get
\[\begin{split}
  - H^{\epsilon+3\lambda}_{\max}(\widetilde{A}F^n|{E'}^n)
                &=    H^{\sin\alpha}_{\min}(\widetilde{A}F^n|E^n)      \\
                &\leq H^{\sin\alpha}_{\max}(\widetilde{A}F^n|E^n) 
                                       + \log\frac{1}{\cos^2(2\alpha)} \\
                &=    H^{\sin\alpha}_{\max}(\widetilde{A}F^n|{E'}^n) 
                                       + 2\log\frac{1}{\cos(2\alpha)},
\end{split}\]
using the symmetry of the pure state $\psi$ with respect to swapping $E^n$ and ${E'}^n$,
as expressed in Lemma~\ref{lemma:degrading}. We find that
\begin{equation}\begin{split}
  \label{eq:symmetric-term}
  - H^{\epsilon+3\lambda}_{\max}(\widetilde{A}F^n|{E'}^n)
                &\leq \log\frac{1}{1-2(\epsilon+3\lambda)^2}            \\
                &\!\!\!
                 =    \log\frac{1}{1-2\left(\threshold-\lambda\right)^2} 
                 \leq \log\frac{1}{2\lambda}.
\end{split}\end{equation}

Turning to the first term in Eq.~(\ref{eq:still-general}), we note that it
is evaluated on $\psi^{F^n{E'}^n} = V^{\ox n}\cN^{\ox n}\bigl(\rho^{(n)}\bigr) V^{\dagger\ox n}$,
a linear function of the input density 
$\rho^{(n)} = \tr_{\widetilde{A}}\phi \in \cS({A'}^n)$. By slight abuse of
notation we henceforth write
\[
  H^\lambda_{\max}(F^n|{E'}^n)_{\rho^{(n)}} = H^\lambda_{\max}(F^n|{E'}^n)_\psi.
\]
Now, if we knew that the maximum of this max-entropy is attained on a tensor power state 
$\rho^{(n)} = \rho^{\ox n}$, then we would be done, by immeditately applying
the asymptotic equipartition property (AEP) for min- and max-entropies 
(Proposition~\ref{prop:AEP}). 
A priori, however, the state $\rho^{(n)}$ is arbitrary
(note that it eventually comes directly from the optimal code with which
we started our reasoning), so we need to work a little more. 
To this end we shall exploit the permutation covariance of the channel; 
for any permutation $\pi\in S_n$, acting naturally on an $n$-partite system, we have
\[
  \pi \psi^{(FE')^n} \pi^\dagger 
      = V^{\ox n}\cN^{\ox n}\bigl(\pi\rho^{(n)}\pi^\dagger\bigr) V^{\dagger\ox n},
\]
and since $\pi^{(FE')^n} = \pi^{F^n} \ox \pi^{{E'}^n}$ and by the local unitary
invariance of the min- and max-entropies, we get
\[
  H^\lambda_{\max}(F^n|{E'}^n)_{\rho^{(n)}} = H^\lambda_{\max}(F^n|{E'}^n)_{\pi\rho^{(n)}\pi^\dagger}.
\]
At this point we can use a restricted concavity property of the max-entropy,
Lemma~\ref{lemma:concavity} below, and get
\begin{equation}\begin{split}
  \label{eq:symmetrized}
  H^\lambda_{\max}(F^n|{E'}^n)_{\rho^{(n)}} 
       &\leq H^{\lambda/\sqrt{2}}_{\max}(F^n|{E'}^n)_{\overline{\rho}^{(n)}} \\
       &\leq H_{\min}^{1-{\frac18}\lambda^2}(F^n|{E'}^n)_{\overline{\rho}^{(n)}},
\end{split}\end{equation}
for the permutation invariant state
\[
  \overline{\rho}^{(n)} = \frac{1}{n!} \sum_{\pi\in S_n} \pi \rho^{(n)} \pi^\dagger,
\]
where we have also invoked Lemma~\ref{lemma:max-min-inequality},
Eq.~(\ref{MaxMinIneq-simpler}), in the second inequality in (\ref{eq:symmetrized}).

It is well-known that such permutation-invariant states are, in several
meaningful senses, approximated by convex combinations of tensor power states; 
such a statement is known as (finite) de Finetti theorem, and here we use it
in the form of the \emph{Post-Selection Lemma}~\cite{CKR-post-select} 
(Lemma~\ref{lemma:post-select} below):\footnote{We 
point out that it is 
also possible to do 
this using Renner's 
Exponential de Finetti 
Theorem~\cite{RennerDeFinetti}, 
which requires a little 
more care to employ, 
but yields bounds quite 
similar to the ones 
obtained in the 
following.} 
\[
  \overline{\rho}^{(n)} \leq n^{|A|^2} \omega^{(n)},
\]
where on the right we have the universal de Finetti state
\[
  \omega^{(n)} = \int {\rm d}\sigma\, \sigma^{\ox n},
\]
for a certain universal measure on states $\sigma \in \cS(A)$.
Without loss of generality, by Carath\'eodory's Theorem, it may be assumed to be 
supported on $M \leq n^{2|A|^2}$ points, hence we may write
\[
  \omega^{(n)} = \sum_{i=1}^M p_i \sigma_i^{\ox n}.
\]
Now we claim that
\begin{equation}
  \label{eq:post-selected}
  H_{\min}^{1-{\frac18}\lambda^2}(F^n|{E'}^n)_{\overline{\rho}^{(n)}}
     \leq H_{\min}^{1-\frac{1}{16}\lambda^2 n^{-|A|^2}}\!\!(F^n|{E'}^n)_{\omega^{(n)}}.
\end{equation}
Indeed, let $\rho'$ be such that 
$P(\rho',\overline\rho^{(n)}) \leq 1-\delta := 1-{\frac18}\lambda^2$. I.e., by the
post-selection inequality and the operator monotonicity of the square root,
\[\begin{split}
  \sqrt{1-(1-\delta)^2}  \leq F(\rho',\overline{\rho}^{(n)})
                        &=    \left\| \sqrt{\rho'}\sqrt{\overline{\rho}^{(n)}} \right\|_1 \\
                        &\leq n^{\frac12 |A|^2} \left\| \sqrt{\rho'}\sqrt{\omega^{(n)}} \right\|_1,
\end{split}\]
thus
\[\begin{split}
  F(\rho',\omega^{(n)})  \geq n^{-\frac12 |A|^2}\sqrt{2\delta-\delta^2} 
                        &\geq \sqrt{\delta n^{-|A|^2}} \\
                        &\geq \sqrt{1-(1-\delta')^2},
\end{split}\]
with $\delta' = \frac12 \delta n^{-|A|^2}$.
Hence, from Eqs.~(\ref{eq:symmetrized}) and (\ref{eq:post-selected}),
Lemma~\ref{lemma:min-max-inequality}, Eq.~(\ref{MinMaxIneq-simpler}), 
and Lemma~\ref{lemma:max+log-bound} below (with the finite-support decomposition 
of $\omega^{(n)}$),
\begin{equation}\begin{split}
  \label{eq:second-term}
  H_{\max}^\lambda(F^n|{E'}^n)_{\rho^{(n)}}
     &\leq H_{\max}^{\frac{1}{32}\lambda^2 n^{-|A|^2}}\!\!(F^n|{E'}^n)_{\omega^{(n)}} 
              + \log \frac{32 n^{|A|^2}}{\lambda^2} \\
     &\leq \max_{\rho\in\cS(A)} H_{\max}^{\frac{1}{32}\lambda^2 n^{-|A|^2}}\!\!(F^n|{E'}^n)_{\rho^{\ox n}} \\
     &\phantom{======}         + 3|A|^2 \log n + 6 + \log\frac{1}{\lambda^2}.
\end{split}\end{equation}

Putting Eqs.~(\ref{eq:still-general}), (\ref{eq:symmetric-term})
and (\ref{eq:second-term}) together, we arrive at
\[\begin{split}
  \log N_E(n,\epsilon) 
     &\leq \max_{\rho\in\cS(A)} H_{\max}^{\frac{1}{32}\lambda^2 n^{-|A|^2}}(F^n|{E'}^n)_{\rho^{\ox n}} \\
     &\phantom{======}         + 3|A|^2 \log n + 5 + 5\log\frac{1}{\lambda}.
\end{split}\]

Note that the optimization over $\rho$ is indeed a maximum since the smooth
max-entropy is a continuous function of the state.
The last step of the proof is an appeal to the quantum asymptotic equipartition
property (Proposition~\ref{prop:AEP}),
\[
  H_{\max}^{\frac{1}{32}\lambda^2 n^{-|A|^2}}(F^n|{E'}^n)_{\rho^{\ox n}} 
       \leq n S(F|E')_{\rho} + \mu\sqrt{n\ln\frac{64 n^{|A|^2}}{\lambda^2}},
\]
and we are done.
\end{IEEEproof}
\medskip

\begin{remark}
The error $\threshold$ is precisely that achieved asymptotically by
a single 50\%-50\% erasure channel acting on the code space, and of other
suitable symmetric (i.e., degradable and anti-degradable) channels. 
We draw attention to the fact that in the proof we encounter a
symmetric state, up to a local unitary, $\psi^{\widetilde{A}F^n:E^n:{E'}^n}$,
which can indeed be interpreted as the joint state between input 
($\widetilde{A}F^n$), output ($E^n$) and environment (${E'}^n$) 
of a suitable test state with a symmetric channel's Stinespring 
dilation.

We need to bound its min-entropy, 
$H_{\min}^{\epsilon+3\lambda}(\widetilde{A}F^n|E^n)$, but if 
$\epsilon \geq \threshold$, then the overall smoothing parameter is strictly larger
than that, and without any additional structure of the state we cannot upper
bound the quantity further: 
Indeed, note that the symmetry we were using is 
consistent with an arbitrarily large entangled state passing through a 
single 50\%-50\% erasure channel of sufficiently large input dimension,
so
\[
  \ket{\psi}^{\widetilde{A}EE'} = \frac{1}{\sqrt{2}}\ket{\Phi}^{\widetilde{A}E}\ket{*}^{E'}
                                  + \frac{1}{\sqrt{2}}\ket{\Phi}^{\widetilde{A}E'}\ket{*}^{E}.
\]
The smoothing by more than $\threshold$ allows us to get rid of the
erasure output on $E$ and pick out the successful generation of a
maximally entangled state, yielding an arbitrarily large smooth min-entropy.

However, in Sections~\ref{sec:from-symmetric-to-degradable} and \ref{sec:SDP}
we will discuss other potential approaches, which might work because they
use all the available structure.
\end{remark}

\medskip
Here are the lemmas needed in the above proof; they are proved in the appendix.

\medskip
It is known that the max-entropy $H_{\max}(A|B)_{\rho}$ is concave in the state 
$\rho_{AB}$ \cite{TCR10}, but this does not extend to the smoothed version. 
However, the following statement holds.

\begin{lemma}
\label{lemma:concavity}
Let $\rho \in \cS(AB)$ be a state and consider the state family
$\rho_i^{AB} = (U_i\ox V_i)\rho(U_i\ox V_i)^\dagger$, with unitaries $U_i$ on $A$
and $V_i$ on $B$, and probabilities $p_i$; define
$\overline{\rho} := \sum_i p_i \rho_i$. Then,
\[
  H_{\max}^{\epsilon}(A|B)_{\overline{\rho}} \geq H_{\max}^{\epsilon\sqrt{2}}(A|B)_{\rho}.
\]
\end{lemma}

\begin{lemma}
  \label{lemma:max+log-bound}
  For an ensemble $\{p_i,\rho_i\}_{i=1}^M$ of states $\rho_i \in \cS(AB)$ with
  probabilities $p_i$, let $\overline{\rho} = \sum_i p_i\rho_i$. Then, for any
  $0\leq \epsilon\leq 1$,
  \[
    H_{\max}^\epsilon(A|B)_{\overline{\rho}} \leq \max_i \ H_{\max}^\epsilon(A|B)_{\rho_i} + \log M.
  \]
\end{lemma}

\begin{lemma}[Post-Selection Technique~\cite{CKR-post-select}]
\label{lemma:post-select}
For a Hilbert space $\cH$ of dimension $d$, denote by $\operatorname{Sym}^n(\cH)$
the subspace of permutation-invariant states in $\cH^{\ox n}$. Then, for
every state $\rho$ supported on $\operatorname{Sym}^n(\cH)$,
\[
  \rho \leq n^d \int {\rm d}\psi\, \proj{\psi}^{\ox n} = P_{\operatorname{Sym}^n(\cH)},
\]
with the uniform (i.e., unitarily invariant) probability measure ${\rm d}\psi$
on pure states of $\cH$, and -- by Schur's Lemma -- the projector
$P_{\operatorname{Sym}^n(\cH)}$ onto the symmetric subspace.

If $\rho$ is a state on $\cH^{\ox n}$ invariant under conjugation by 
permutations, $\rho = \pi\rho\pi^\dagger$ for all $\pi\in S_n$, then the above
can be applied to its purification in $\operatorname{Sym}^n(\cH\ox\cH')$,
giving
\[
  \rho \leq n^{d^2} \int {\rm d}\sigma\, \sigma^{\ox n},
\]
with a universal probability measure ${\rm d}\sigma$ on $\cS(\cH)$.
\altqed
\end{lemma}

Finally, we state a simplified version of the asymptotic equipartition property
for min- and max-entropies, giving useful bounds for every $n$: 
\begin{proposition}[Min- and max-entropy AEP \cite{Renner:PhD,TomamichelThesis}]
\label{prop:AEP}
Let $\rho \in \mathcal{S}(\mathcal{H}_{AB})$ and $0 < \epsilon < 1$. 
Then, 
\[\begin{split}
\lim_{n \to \infty} \frac{1}{n} H^{\epsilon}_{\min}(A^n | B^n)_{\rho^{\ox n}} 
                    &= \lim_{n \to \infty} \frac{1}{n} H^{\epsilon}_{\max}(A^n | B^n)_{\rho^{\ox n}} \\
                    &= S(A|B)_{\rho}.
\end{split}\]

More precisely, for a purification $\ket{\psi} \in ABC$ of $\rho$, denote 
$\mu_X := \log \left\|(\psi^X)^{-1}\right\|$, where the inverse is the generalized
inverse (restricted to the support), for $X=B,C$. Then, for every $n$, 
\begin{align}
  \label{eq:H_min_lower}
  H_{\min}^\epsilon(A^n|B^n) &\geq n S(A|B) - (\mu_B+\mu_C)\sqrt{n\ln\frac{2}{\epsilon}}, \\
  \label{eq:H_max_upper}
  H_{\max}^\epsilon(A^n|B^n) &\leq n S(A|B) + (\mu_B+\mu_C)\sqrt{n\ln\frac{2}{\epsilon}},
\end{align}
and similar opposite bounds via Lemma~\ref{lemma:min-max-inequality}.
\end{proposition}

\section{\Semistrong{} converse\protect\\ for the private capacity}
\label{sec:P-semi-strong}
In this section we show that the argument in the previous section can be augmented 
to yield a pretty strong converse for the private capacity.

We start by reviewing the basic definitions, which we adapt from Renes and
Renner~\cite{RenesRenner}: 
A \emph{private classical code} for a channel $\cN:\cL(A') \rightarrow \cL(B)$ 
consists of a family of signal states $\rho_x \in \cS(A')$ ($x=1,\ldots,M$),
and a decoding measurement (POVM) $(D_x)_{x=1}^M$, i.e.~$D_x \geq 0$, $\sum_x D_x = \1_B$.
The latter can also be viewed as a cptp map $\cD:\cL(B) \rightarrow \widehat{X}$.
Postulating a uniform distribution on the messages $x$, the code gives rise to the
following averaged ccq-state of input, output and environment:
\[
  \sigma^{X\widehat{X}E} = \frac{1}{M} \sum_x \proj{x}^X \ox \proj{\hat{x}}^{\widehat{X}} 
                                                         \ox \tr_B (V\rho_x V^\dagger)(D_{\hat{x}} \ox \1_E),
\]
encoding all correlations between legal users and eavesdropper of the system.
The \emph{error} of the code is defined in terms of the purified distance as
\[\begin{split}
  P&\left( \frac{1}{M}\sum_x \proj{x}^X \ox \proj{x}^{\widehat{X}}, \sigma^{X\widehat{X}} \right) \\
   &\phantom{======}
    = \sqrt{1-\left( \frac{1}{M}\sum_x \sqrt{\tr \cN(\rho_x)D_x} \right)^2}.
\end{split}\]
Its \emph{privacy} is defined as
\[\begin{split}
  \min_{\widetilde{\rho} \in \cS(E)} 
     P &\left( \frac{1}{M}\sum_x \proj{x}^X \ox \widetilde{\rho}^E, \sigma^{XE} \right) \\
       &\phantom{====}
        = \min_{\widetilde{\rho} \in \cS(E)}
            \sqrt{1-\left( \frac{1}{M}\sum_x F(\cN^c(\rho_x),\widetilde{\rho}^E) \right)^2}.
\end{split}\]
For a given channel $\cN$, we denote the largest $M$ such that there exists 
a private classical code with error $\epsilon$ and privacy $\delta$,
by $M(n,\epsilon,\delta)$. The \emph{(weak) private capacity} of $\cN$ is then
defined as
\[
  P(\cN) = \inf_{\epsilon,\delta > 0}
                 \liminf_{n\rightarrow\infty} \frac{1}{n} \log M(n,\epsilon,\delta).
\]
It was determined in~\cite{Devetak,CaiWinterYeung}, and like $Q$ it is
only known as a regularized characterization in general \cite{SmithRenesSmolin08}. 
By the monogamy of entanglement, we know that $P(\cN) \geq Q(\cN)$ 
(see the Remark below), but in general this inequality is strict.

However for degradable channels, it was proved by Smith~\cite{Smith08} that
the private capacity $P(\cN)$ equals the quantum capacity $Q(\cN)=Q^{(1)}(\cN)$,
and is hence given by a simple single-letter formula.

\medskip
\begin{remark}
The way we defined the code and the error above (as an average) is really that
of a secret key generation code, analogous to the entanglement-generating
codes in the previous section. 

This (long) remark is about an alternative
definition with worst case errors and privacy over individual messages.
Indeed, such a notion is stronger and will imply error and privacy as
we defined them above. To go conversely from averaged error and privacy  
to essentially the same worst-case notions at 
the expense of loosing a constant fraction of the messages (hence no rate
loss asymptotically) we use Ahlswede's observation \cite{Ahlswede78} 
on how randomization in the encoding can turn \emph{several} average 
errors into only slightly worse worst-case errors. 

For a code with messages $x=1,\dots,M$ and joint cq-state after decoding,
\[
  \rho^{ABE} = \frac1M \sum_{xy} P(y|x) \proj{x} \otimes \proj{y} \otimes \rho_{xy}^E,
\]
consider the reduced states
\begin{align*}
  \rho^{AB} &= \frac1M \sum_{xy} P(y|x) \proj{x} \otimes \proj{y}, \\
  \rho^{AE} &= \frac1M \sum_{x}  \proj{x} \otimes \rho_x^E.
\end{align*}
With error and privacy are defined as above,
\[
  \epsilon = P\left(\rho^{AB},  \frac1M  \sum_{x} \proj{x} \otimes \proj{x} \right)
\]
and
\[
  \delta = P\left(\rho^{AE},  \frac1M  \sum_{x} \proj{x} \otimes \sigma^E \right),
\]
where $P=\sqrt{1-F^2}$ is the purified distance, a short calculation shows that
\begin{align*}
  F\left(\rho^{AB},  \frac1M \sum_{x} \proj{x} \otimes \proj{x}\right) 
                                        &=  \frac1M  \sum_x \sqrt{P(x|x)} =: F_1,        \\
  F\left(\rho^{AE},  \frac1M \sum_{x} \proj{x} \otimes \sigma^E\right) 
                                        &=  \frac1M  \sum_x F(\rho_x^E,\sigma^E) =: F_2.
\end{align*}
We will now encode messages $m$ into uniform distributions on pairwise disjoint
sets $K_m \subset [M] = \{1,\ldots,M\}$ 
of cardinality $k$, with $m=1,\ldots, N$ such that $kN \leq M$. 

We will draw the elements of $K_1,\dots,K_N$ randomly and without replacement from $[M]$.
We then use Azuma's inequality to bound the probability that for a given $m$ and 
$\eta \geq 0$
\[
  \frac1k \sum_{x \in K_m} \sqrt{P(x|x)} < F_1 - \eta, 
\]
or
\[
  \frac1k \sum_{x \in K_m} F(\rho_x^E,\sigma^E) < F_2 - \eta.
\]
Namely, each of these events has probability at most 
$p = 2 e^{-2k \eta^2}$ \cite{Azuma,DemboZeitouni}. 
The input-output-environment state of the new code for the messages $m=1,\dots,N$ is 
\[
  \omega = \frac{1}{N} \sum_{mm'} \frac1k \sum_{x \in K_m, y\in K_{m'}} 
                         P(y|x) \proj{m} \otimes \proj{m'} \otimes \rho_x^E.
\]
Note that $P(m|m) \geq \frac1k \sum_{x \in K_m} P(x|x)$, and by concavity of the 
square root,
\[
  \sqrt{P(m|m)} \geq \frac1k \sum_{x \in K_m} \sqrt{P(x|x)}.
\]

Likewise, the state of the eavesdropper for message $m$ is $\frac1k \sum_{x \in K_m} \rho_x^E$,
and by concavity of the fidelity,
\[
  F\left(\frac1k \sum_{x \in K_m} \rho_x^E, \sigma^E \right) 
                               \geq \frac1k \sum_{x \in K_m} F(\rho_x^E, \sigma^E).
\]
I.e., this message will have individual error $\leq \epsilon'$ and 
individual privacy $\leq \delta'$
for these ``good'' $m$, where it is straightforward to work out that
$\epsilon' \leq \epsilon \left( 1 + \frac{\eta}{\epsilon^2} \right)$ and
$\delta'   \leq \delta   \left( 1 + \frac{\eta}{\delta^2} \right)$.
In other words, by choosing $\eta = a \cdot \min(\epsilon^2,\delta^2)$ we can make the new error
and privacy arbitrarily close to the original parameters.

Now, we can find $K_1,\ldots,K_N$ such that a fraction $\geq 1-p$ of the $K_m$ are
``good'', throw away the ``bad'' $m$ and we are left with the code we want:
it has $N' \geq (1-p)N = \frac{1-p}{k} M \geq \frac{1}{2k} M$ messages, 
if we choose $k$ such that $p \leq 1/2$,
which holds for~$k \geq \frac{\ln 4}{2\eta^2}$. 

In summary, we can get a code with randomized encoding and individual error 
$\epsilon' < (1+a)\epsilon$ and individual privacy $\delta' < (1+a)\delta$
for each message, and losing a constant amount of information compared to
the original code we started from. Indeed the number of bits encoded
diminishes by at most
\[
  2 \log \frac{1}{\eta} \leq 2 \log \frac1a + 4 \log \frac{1}{\epsilon} + 4 \log \frac{1}{\delta}.
\]
\end{remark}

\medskip
By definition, every entanglement-generating code of error $\epsilon'$ gives rise to a 
private classical (secret key generation) code of error
and privacy $\epsilon'$, and with $M=|C|$ messages. Thus,
$M(n,\epsilon',\epsilon') \geq N_E(n,\epsilon') \geq N(n,\epsilon')$.

\medskip
\begin{theorem}
  \label{thm:P-semi-strong}
  Let $\cN:\cL(A) \rightarrow \cL(B)$ be a degradable channel with finite
  quantum systems $A$ and $B$. Then, for error $\epsilon$ and
  privacy $\delta$ such that $\epsilon + 2\delta < \threshold$
  (e.g. $\epsilon = \delta < \frac{1}{3\sqrt{2}} \approx .2357$),
  and every integer $n$,
  \[\begin{split}
    \log M(n,\epsilon,\delta) &\leq n Q^{(1)}(\cN) + \mu\sqrt{n\ln\frac{64 n^{|A|^2}}{\eta^2}} \\
                              &\phantom{\leq n Q^{(1)}(\cN)}
                               + 3|A|^2 \log n + 9 + 11\log\frac{1}{\eta},                     \\
                              &\leq n Q^{(1)}(\cN) + O\left(\sqrt{n\log n}\right),    
  \end{split}\]
  where $\eta = \frac16\left(\threshold-\epsilon-2\delta\right)$.
\end{theorem}

Together with the direct part (achievability proved in~\cite{Devetak,CaiWinterYeung})
we thus get:
\begin{corollary}
For a degradable channel $\cN$, the private capacity is given by
\[
  P(\cN) = \lim_{n\rightarrow\infty} \frac{1}{n}\log M(n,\epsilon,\delta),
\]
for any $\epsilon,\delta > 0$ such that $\epsilon + 2\delta < \threshold$.
\altqed
\end{corollary}

\medskip
\begin{IEEEproof}
Consider a code for $\cN^{\ox n}$ with $M=M(n,\epsilon,\delta)$ messages,
that has error $\epsilon$ and is $\delta$-private: message $x$ (chosen 
uniformly) is encoded as $\sigma_x \in \cS({A'}^n)$ and sent through the 
channel, giving rise to an averaged cqq-state between reference $X$, output $B^n$ 
and environment $E^n$:
\[
  \rho^{XB^nE^n} = \frac{1}{M}\sum_x \proj{x}^X \ox U^{\ox n} \rho_x U^{\dagger\ox n}.
\]
The ``trivial'' converse shows that
\[
  \log M \leq H_{\min}^\delta(X|E^n) - H_{\max}^\epsilon(X|B^n),
\]
cf.~Renes and Renner~\cite{RenesRenner}, whose argument we briefly repeat here 
since they used trace norm rather than purified distance.
According to the definition of privacy given above, the reduced state $\rho^{XE^n}$
is within purified distance $\delta$ of a product state of the form 
$\frac{1}{M}\sum_x \proj{x} \ox \widetilde{\rho}^{E^n}$, hence 
$H_{\min}^\delta(X|E^n) \geq \log M$. Likewise, there exists a decoding cptp
map $\cD:\cL(B^n) \rightarrow \widehat{X}$ such that $(\id\ox\cD)\rho^{XB^n}$ 
is within $\epsilon$ purified distance from the perfectly correlated state
$\frac{1}{M}\sum_x \proj{x}^X \ox \proj{x}^{\widehat{X}}$, hence
$H_{\max}^\epsilon(X|B^n) \leq 0$. 

Now we can purify $\rho^{XB^nE^n} = \tr_{A_0X'} \varphi^{XX'A_0B^nE^n}$, 
introducing a dummy system $A_0$ to hold the purifications $\phi_x^{A_0{A'}^n}$ 
of the signal states $\rho_x$ and a coherent copy $X'$ of $X$:
\[
  \ket{\varphi}^{XX'B^nE^n}
     = \frac{1}{\sqrt{M}}\sum_x \ket{x}^X\ket{x}^{X'}(\1_{XX'} \ox U^{\ox n})\ket{\phi_x}^{A_0{A'}^n},
\]
to which we then also apply the Stinespring dilation of the degrading map:
\[\begin{split}
  \ket{\psi}^{XX'{E'}^nF^nE^n}
     &= (\1_{XX'} \ox V^{\ox n} \ox \1_{E^n})\ket{\varphi} \\
     &\!\!\!\!\!\!\!\!\!\!\!\!\!\!\!\!\!\!\!\!\!
      = \frac{1}{\sqrt{M}}\sum_x \ket{x}^X\ket{x}^{X'}
                                  \bigl(\1_{XX'} \ox (VU)^{\ox n}\bigr)\ket{\phi_x}^{A_0{A'}^n},
\end{split}\]

With respect to $\psi$, we thus have
\begin{equation}\begin{split}
  \label{eq:privacy-bound}
  \log M &\leq H_{\min}^\delta(X|E^n) - H_{\max}^\epsilon(X|{E'}^nF^n) \\
         &=    H_{\min}^\delta(X|{E'}^n) - H_{\max}^\epsilon(X|{E'}^nF^n) \\
         &\leq H_{\max}^\eta(F^n|{E'}^n) - H_{\max}^{\epsilon+2\delta+5\eta}(F^n|{E'}^nX) \\
         &\phantom{=================}
               + 4\log\frac{2}{\eta^2},
\end{split}\end{equation}
where we have used the degradability property of the channel in the second line,
and in the third line the chain rule, Lemma~\ref{lemma:chainRule}, in its two manifestations
Eqs.~(\ref{eq:chain:max-le-max+max}) and (\ref{eq:chain:max-ge-min+max}). 
Indeed,
\begin{align*}
  H_{\max}^{\epsilon+3\eta}(AB|C) &\leq H_{\max}^\eta(A|C) + H_{\max}^\epsilon(B|AC) 
                                                                               + \log\frac{2}{\eta^2} \\
    \parallel \phantom{===}        &        \\
  H_{\max}^{\kappa}(AB|C)         &\geq H_{\min}^{\delta}(B|C) + H_{\max}^{\kappa+2\delta+2\eta}(A|BC)  \\
  &\phantom{=================}
                                                                              - 3\log\frac{2}{\eta^2},
\end{align*}
which we employ with the identifications $F^n \equiv A$, $X\equiv B$, ${E'}^n \equiv C$,
and with $\kappa = \epsilon+3\delta$.

Choosing $\eta = \frac16\left(\threshold-\epsilon-2\delta\right)$ ensures that
$\epsilon' := \epsilon+2\delta+5\eta = \threshold - \eta < \threshold$, and we can bound
the second term on the right hand side of Eq.~(\ref{eq:privacy-bound}) as
before, in the proof of Theorem~\ref{thm:Q-semi-strong}:
\[\begin{split}
  -H_{\max}^{\epsilon'}(F^n|{E'}^nX) 
      &\leq -H_{\min}^{\epsilon'}(F^n|{E'}^nX) + 2\log\frac{1}{2\eta} \\
      &=     H_{\max}^{\epsilon'}(F^n|E^nX'A_0) + 2\log\frac{1}{2\eta} \\
      &\leq  H_{\max}^{\epsilon'}(F^n|E^nX') + 2\log\frac{1}{2\eta} \\
      &=     H_{\max}^{\epsilon'}(F^n|{E'}^nX) + 2\log\frac{1}{2\eta},
\end{split}\]
where we have used Lemma~\ref{lemma:min-max-inequality}, 
then the duality between min- and max-entropy, then
the monotonicity (Lemma~\ref{lemma:mono}) and finally the exchange symmetry between $X$ and $X'$
as well as between $E$ and $E'$. As this means
\[
  -H_{\max}^{\epsilon'}(F^n|{E'}^nX) \leq \log\frac{1}{2\eta},
\]
we have by plugging this into Eq.~(\ref{eq:privacy-bound}),
\[
  \log M(n,\epsilon,\delta) \leq H_{\max}^\eta(F^n|{E'}^n) + 3 + 9\log\frac{1}{\eta},
\]
and the rest of the argument is as in the proof of Theorem~\ref{thm:Q-semi-strong}
[cf.~Eq.~(\ref{eq:second-term})]:
\[\begin{split}
  H_{\max}^\eta(F^n|{E'}^n)
     &\leq H_{\max}^{\frac{1}{32}\eta^2 n^{-|A|^2}}\!\!(F^n|{E'}^n)_{\omega^{(n)}} 
              + \log \frac{32 n^{|A|^2}}{\eta^2} \\
     &\leq \max_{\rho\in\cS(A)} H_{\max}^{\frac{1}{32}\eta^2 n^{-|A|^2}}\!\!(F^n|{E'}^n)_{\rho^{\ox n}} \\
     &\phantom{======}         + 3|A|^2 \log n + 6 + \log\frac{1}{\eta^2},
\end{split}\]
invoking the quantum AEP for the max-entropy (Proposition~\ref{prop:AEP}).
\end{IEEEproof}

\section{Strong converse for symmetric channels implies it for degradable channels}
\label{sec:from-symmetric-to-degradable}
The main result of this section, Theorem~\ref{thm:Q-strong:from-symmetric-to-degradable},
is valid for degradable channels satisfying the following technical condition.
\begin{definition}
  \label{defi:type-I}
  We say that a degradable channel $\cN$ is of \emph{type I} (for \emph{invariance})
  if one can choose a Stinespring dilation $U$ of it, and a Stinespring
  dilation $V$ of a degrading channel $\cM$, such that the unitary $X_F$ in
  Lemma~\ref{lemma:degrading} is a global phase (hence $\pm 1$). I.e.,
  \[
    (\1_F \ox \SWAP_{EE'})UV = \pm UV.
  \]
\end{definition}

\begin{example}
  \emph{(Erasure channels).} The qubit erasure channel
  \[
     \cE_q(\rho) = (1-q)\rho \oplus q\proj{\ast}
  \]
  with erasure probability $q\leq\frac12$ has as its
  complementary channel $\cE_q^c = \cE_{1-q}$; as degrading map
  serves $\cE_t$, with $t=\frac{q}{1-q}$ (augmented by the identity on $\proj{\ast}$).
  
  We can guess an isometric dilation of $\cE_q$,
  \[
    U:\ket{\phi} \longmapsto \sqrt{1-q}\ket{\phi}^B\ket{\ast}^E 
                                 + e^{i\alpha}\sqrt{q}\ket{\ast}^B\ket{\phi}^E,
  \]
  and likewise for the degrading map,
  \begin{align*}
    V:\ket{\ast} &\longmapsto \ket{\ast}^F\ket{\ast}^{E'}\\
      \ket{\phi} &\longmapsto \sqrt{1-t}\ket{\phi}^F\ket{\ast}^{E'} + \sqrt{t}\ket{\ast}^F\ket{\phi}^{E'}.
  \end{align*}
  With the choice of phase $e^{i\alpha} = 1$, it is straightforward to verify that 
  $\SWAP_{EE'}VU = VU$.
   
  However, since the output of an erasure channel has no coherences between
  the erasure symbol and the unerased part, there is considerable freedom
  in choosing the dilations both of the channel and of the degrading map.
  For some of them there is no unitary $X_F$ as in Lemma~\ref{lemma:degrading},
  for some the unitary is non-trivial. Indeed, we can see this by varying
  $\alpha$ in the dilation $U$ above, most choices of which leave no symmetry
  $X_F$, but for $e^{i\alpha}=-1$ we can choose $X_F=2\proj{\ast}-\1$.
\end{example}

\medskip
\begin{example}
  \emph{(Schur multiplier channels).} Given a positive semidefinite
  $n\times n$-matrix $S\geq 0$ with diagonal entries $S_{ii}=1$ one can define a 
  cptp map $\cN_S$ on $n\times n$-matrices by Schur/Hadamard multiplication of the 
  input $\rho$ by $S$:
  \[
    \cN_S :\rho             \longmapsto \rho \circ S, \text{ i.e. } 
    \cN_S(\ket{i}\!\bra{j}) = S_{ij} \ket{i}\!\bra{j}.
  \]
  
  It is well-known that $S$ can be viewed as Gram matrix of unit vectors 
  $\ket{\varphi_1},\ldots,\ket{\varphi_n}$:
  \[
    S_{ij} = \bra{\varphi_j}\varphi_i\rangle,
  \]
  suggesting a Stinespring dilation
  \[
    U:\ket{i} \longmapsto \ket{i}^B\ket{\varphi_i}^E.
  \]
  It gives rise to the complementary channel 
  \[
    \cN_S^c(\ket{i}\!\bra{j}) = \delta_{ij} \proj{\varphi_i},
  \]
  so we can choose $\cN_S^c$ itself as degrading map and essentially $U$ as
  its dilation $V$ (with $F$ taking the place of $B$, and $E'$ that of $E$).
  
  Thus,
  \[
    VU:\ket{i} \longmapsto \ket{i}^F\ket{\varphi_i}^E\ket{\varphi_i}^{E'},
  \]
  which is evidently invariant under $\SWAP_{EE'}$ since the output 
  state restricted to $EE'$, $\tr_F VU \rho U^\dagger V^\dagger$, is
  supported on the symmetric subspace of $E\ox E'$.
\end{example}

\medskip
\begin{remark}
We do not know whether all degradable channels are of type I, not having found a 
counterexample so far. From the examples given above it is clear however that 
the dilations $U$ and $V$ required for a proof that a given channel is type I, 
have to be constructed carefully. The next lemma shows that for any degradable
channel we can construct one that is information theoretically equivalent,
and which is of type I.
\end{remark}

\medskip
\begin{lemma}
  \label{lemma:degrading-type-I}
  For every degradable channel $\cN:\cL(A') \rightarrow \cL(B)$, the channel
  \begin{align*}
    \widetilde{\cN} = \cN \ox \tau^{B_0} : \cL(A') &\rightarrow \cL(B\ox B_0), \\
                                           \rho    &\mapsto \cN(\rho) \ox \tau^{B_0},
  \end{align*}
  which attaches to the output of $\cN$ a qubit system $B_0$ in the maximally
  mixed state, is degradable of type I.
\end{lemma}
\begin{IEEEproof}
Clearly, $\widetilde{\cN}^c = \cN^c \ox \tau^{E_0}$, with a qubit system
$E_0$, so the new channel is also degradable.

Choose a Stinespring isometry $U$ of $\cN$ and $V$ of the degrading map $\cM$
according to Lemma~\ref{lemma:degrading}, so that we have a unitary involution
$X_F$ with
\[
  (X_F\ox\SWAP_{EE'})VU = VU.
\]
$X_F$ can have only the two eigenvalues $\pm 1$, so decompose $F = F_+ \oplus F_-$
into the respective eigenspaces with projectors $P_+$ and $P_-$, respectively.
Of course also $\SWAP_{EE'}$ has eigenvalues $\pm 1$, the corresponding eigenspaces
being known as symmetric and anti-symmetric subspace, denoted as $\Sym^2(E)$
and $\Lambda^2(E)$, respectively.

The above invariance of $VU$ under left multiplication by $X_F\ox\SWAP_{EE'}$
is equivalently expressed by saying that $VU$ maps $A'$ into the
$+1$-eigenspace of $X_F\ox\SWAP_{EE'}$, which is
\[
  F_+ \ox \Sym^2(E) \oplus F_- \ox \Lambda^2(E).
\]

In this picture we see why $X_F$ is necessary: it is there to
undo a possible phase of $-1$ induced by $\SWAP_{EE'}$ (on $\Lambda^2(E)$),
by applying the same phase once more on $F_-$. We can also see how to
write down dilations of $\widetilde{\cN}$ and a degrading map that avoid this
problem: First, $\widetilde{U}:A' \hookrightarrow (B\ox B_0) \ox (E\ox E_0)$
with
\[
  \widetilde{U}\ket{\phi} 
     := (U\ket{\phi})^{BE} \ox \left( \frac{\ket{01}+\ket{10}}{\sqrt{2}} \right)^{B_0E_0}
\]
is a dilation of $\widetilde{\cN}$. Secondly, we define a degrading map
by writing down directly an isometric dilation 
$\widetilde{V}:B\ox B_0 \hookrightarrow F \ox (E'\ox E_0')$:
\[
  \widetilde{V}(\ket{\varphi}^B\ket{b}^{B_0})
     := \left(\1_{E'} \ox \operatorname{C-Z}^{F\rightarrow E_0'}\right)\bigl((V\ket{\varphi})\ket{b}\bigr),
\]
where
\[
  \operatorname{C-Z}^{F\rightarrow E_0'} = P_+ \ox \1_{B_0} + P_- \ox Z_{B_0}
\]
is a controlled-Z using the $F_{\pm}$ subspaces to trigger a $Z$ on the
qubit $E_0'$ (which we identify with $B_0$).

It is easy to check that $\tr_F \widetilde{V}\cdot\widetilde{V}^\dagger$ defines
a bone fide degrading map for $\widetilde{\cN}$. But it is also of type I,
as it can be confirmed by direct calculation that
\[\begin{split}
  \widetilde{V}\widetilde{U}\ket{\phi}
     &= (P_+\ox\1_{EE'})VU\ket{\phi} \ox \left( \frac{\ket{01}+\ket{10}}{\sqrt{2}} \right)^{E_0E_0'} \\
     &\phantom{==}
       + (P_-\ox\1_{EE'})VU\ket{\phi} \ox \left( \frac{\ket{01}-\ket{10}}{\sqrt{2}} \right)^{E_0E_0'}.
\end{split}\]
Since the left hand factor in the first line is in $F\ox\Sym^2(E)$, while the
analogous term in the second line is in $F\ox\Lambda^2(E)$, the entire expression
lies in $F\ox\Sym^2(EE_0)$, hence under the simultaneous swap $EE_0 \leftrightarrow E'E_0'$,
\[
  \SWAP_{EE_0:E'\!E_0'}\widetilde{V}\widetilde{U} = \widetilde{V}\widetilde{U},
\]
and we are done.
\end{IEEEproof}

\medskip
Degradable channels of type I are intimately related to symmetric channels,
as shown in the next lemma.

\begin{lemma}
  \label{lemma:type-I}
  Let $\cN$ be a degradable channel of type I, and choose a Stinespring 
  dilation $U$ as well as a dilation $V$ of a degrading map, according to
  Lemma~\ref{lemma:degrading}, s.t.~$X_F=\pm\1$. 
  
  For any test state $\ket{\phi_0} \in AA'$ of maximal Schmidt rank, let 
  $\ket{\psi_0}^{AFEE'} = VU\ket{\phi_0}^{AA'}$ and denote the supporting
  subspace of $\psi_0^{AF}$ by $G$.

  Then there is a symmetric channel $\cM$ with Stinepring isometry 
  $W:G'\hookrightarrow E\ox E'$ (i.e.~$\SWAP_{EE'}W = \pm W$) such that every state 
  $\ket{\psi}^{AFEE'} = VU\ket{\phi}$, for $\ket{\phi} \in AA'$ can be
  written as $W\ket{\xi} \in GEE'$ for a suitable test state $\ket{\xi} \in GG'$,
  up to a (state-dependent) isometry $\widehat{W}:G \hookrightarrow AF$:
  \[
    \ket{\psi}^{AFEE'} = (\widehat{W}\ox W)\ket{\xi}^{GG'}.
  \]
\end{lemma}
\begin{IEEEproof}
By definition, $\ket{\psi_0}^{AFEE'} \in G\ox E \ox E'$, so we may denote
it as well $\ket{\psi_0}^{GEE'}$. Choose a purification $\ket{\chi}^{GG'}$
of $\psi_0^G$ with $G' \simeq G$, 
so that there exists an isometry $W:G'\hookrightarrow EE'$ with
\[
  (\1\ox W)\ket{\chi}^{GG'} = \ket{\psi_0}^{GEE'}.
\]
It is easy to see that $W$ has the required symmetry property: since
$\SWAP_{EE'}\ket{\psi_0}^{GEE'} = \pm \ket{\psi_0}^{GEE'}$, it follows
that $(\1\ox \SWAP_{EE'}W)\ket{\chi}^{GG'} = \pm (\1\ox W)\ket{\chi}^{GG'}$,
and since $\ket{\chi}$ has maximal Schmidt rank, $\SWAP_{EE'}W = \pm W$
follows.

Now, let $\ket{\phi}^{AA'}$ be an arbitrary input test state and
$\ket{\psi}^{AFEE'} = VU\ket{\phi}$. Then,
\[
  \ket{\phi}^{AA'} = \left(\sqrt{\phi^A}\sqrt{\phi_0^A}^{-1} \ox \1\right)\ket{\phi_0}^{AA'},
\]
and thus
\[\begin{split}
  \ket{\psi}^{AFEE'} 
        &= \left(\sqrt{\phi^A}\sqrt{\phi_0^A}^{-1} \ox \1^F \ox \1^{EE'}\right)\ket{\psi_0}^{AFEE'} \\
        &= \left(\sqrt{\psi^{AF}}\sqrt{\psi_0^{AF}}^{-1} \ox \1^{EE'}\right)\ket{\psi_0}^{AFEE'} \\
        &= \left(\sqrt{\psi^{AF}}\sqrt{\psi_0^{AF}}^{-1} \ox W\right)\ket{\chi}^{GG'}.
\end{split}\]
Finally, since $\chi^G = \psi_0^{AF}$ has support on $G$, there exists a 
$\sigma \in \cS(G)$ and an isometry $\widehat{W}:G \hookrightarrow AF$
such that
\[\begin{split}
  \left(\sqrt{\psi^{AF}}\sqrt{\phi_0^{AF}}^{-1} \ox \1^{G'}\right)&\ket{\chi}^{GG'}           \\
        &\!\!\!\!\!\!\!\!
         =  \left(\widehat{W}\sqrt{\sigma}\sqrt{\chi_G}^{-1}\ox\1^{G'}\right)\ket{\chi}^{GG'} \\
        &\!\!\!\!\!\!\!\!
         =: (\widehat{W}\ox\1^{G'})\ket{\xi}^{GG'}.
\end{split}\]
In total, $\ket{\psi}^{AFEE'} = (\widehat{W}\ox W)\ket{\xi}^{GG'}$, which is
what we wanted to prove.
\end{IEEEproof}
 
\medskip
\begin{theorem}
\label{thm:Q-strong:from-symmetric-to-degradable}
Let $\cN: \cL(A) \rightarrow \cL(B)$ be a degradable channel, which w.l.o.g.~we
assume to be of type I (by Lemma~\ref{lemma:degrading-type-I}).
Denote its environment by $E$ and the associated symmetric channel by $\cM$, 
with Stinespring dilation $W:G \hookrightarrow E\ox E'$ from Lemma~\ref{lemma:type-I}. 
Then $\cN$ obeys the strong converse for its quantum capacity, if $\cM$ does
(note that by the no-cloning argument, $Q(\cM)=0$). 
More precisely, there exists a constant $\mu$ such that
\[\begin{split}
  \log N_E(n,\epsilon|\cN) 
                 &\leq n Q^{(1)}(\cN) + \mu\sqrt{n\ln\frac{64 n^{|A|^2}}{\lambda^2}} \\
                 &\phantom{\leq n Q^{(1)}(\cN)}
                                      + 8\log\frac{1}{\lambda} + O(\log n)           \\
                 &\phantom{\leq n Q^{(1)}(\cN)}
                                      + \log N_E(n,1-\lambda|\cM),
\end{split}\]
with $\lambda = \frac{1-\epsilon}{5}$.
\end{theorem}
\begin{IEEEproof}
We follow the initial steps of the proof of Theorem~\ref{thm:Q-semi-strong},
until the bound
\[
  \log N_E(n,\epsilon) \leq H_{\max}^\lambda(F^n|{E'}^n)
                            - H_{\max}^{\epsilon+3\lambda}(AF^n|{E'}^n) + \log\frac{2}{\lambda^2},
\]
where all entropies are with respect to the state $\ket{\psi}^{AE^nF^n{E'}^n}$. 
Now we choose $\lambda = \frac{1-\epsilon}{5}$. 

The first term is treated in the exact same way as we did there, giving
\[\begin{split}
  H_{\max}^\lambda(F^n|{E'}^n)
     &\leq \max_{\rho\in\cS(A)} H_{\max}^{\frac{1}{32}\lambda^2 n^{-|A|^2}}\!\!(F^n|{E'}^n)_{\rho^{\ox n}} \\
     &\phantom{======}         + 3|A|^2 \log n + 6 + \log\frac{1}{\lambda^2}                              \\
     &\leq n\, Q^{(1)}(\cN) + \mu\sqrt{n\ln\frac{64 n^{|A|^2}}{\lambda^2}} \\
     &\phantom{======}         + 3|A|^2 \log n + 6 + \log\frac{1}{\lambda^2},
\end{split}\]
where we have used the quantum AEP (Proposition~\ref{prop:AEP}) once more.

The second term can be upper bounded
\[\begin{split}
  - H_{\max}^{\epsilon+3\lambda}(AF^n|{E'}^n)
     &=    H_{\min}^{\epsilon+3\lambda}(AF^n|E^n)_{\psi}                   \\
     &=    H_{\min}^{\epsilon+3\lambda}(G^n|E^n)_{(\1\ox W)^{\ox n}\ket{\xi}}  \\
     &\leq \log N_E(n,\epsilon+4\lambda|\cM) + 4\log\frac{1}{\lambda},
\end{split}\]
using duality in the first equation and Lemma~\ref{lemma:type-I} in the second,
to rewrite the state $\ket{\psi}^{AF^nE^n{E'}^n}$ (up to an isometry 
$G^n \hookrightarrow AF^n$) as if a test state $\ket{\xi}^{G^n{G'}^n}$
had gone through $W^{\ox n}$. The inequality in the third line is by
Proposition~\ref{prop:one-shot-quantum-capacity} below.

Putting these bounds together yields the statement of the theorem.
\end{IEEEproof}

\medskip
The following result is essentially a version of the one-shot
decoupling proof of entanglement-distillation and random quantum coding,
adapted so that the error is composed of a smoothing and a random coding
component; its proof can be found in the appendix.
Note that it gives an essentially matching lower bound to the
upper bound we used in the proof of Theorem~\ref{thm:Q-semi-strong}. It allows
us to assess one of the max-entropy terms we encountered there in a new
light.

\begin{proposition}[Cf.~Buscemi/Datta~\cite{BuscemiDatta} \&{} Datta/Hsieh~\cite{DattaHsieh}]
  \label{prop:one-shot-quantum-capacity}
  Let $U:A' \hookrightarrow B \ox E$ be the Stinespring dilation of a quantum
  channel $\cN$ and $\ket{\phi} \in AA'$ a state vector, 
  $\ket{\psi} := (\1\ox U)\ket{\phi} \in ABC$. Then, given $\eta \geq 0$ and 
  $\epsilon > 0$, there exists an entanglement-generating code for $\cN$,
  creating a maximally entangled state of rank $d$ with error $\leq \eta+\epsilon$,
  where
  \[
    d = \left\lfloor \exp\left( H_{\min}^\eta(A|E)_\psi - 4\log\frac{1}{\epsilon} \right) \right\rfloor.
  \]
\end{proposition}

\medskip
\begin{remark}
  We gave the very precise form of the bounds above to emphasize that if
  the strong converse holds in its exponential form for $\cM$, in the sense
  that for every error rate $c > 0$,
  \[
    \limsup_{n\rightarrow\infty} \frac{1}{n}\log N_E(n,1-2^{-cn}|\cM) \leq f(c),
  \]
  with some non-decreasing continuous function $f(c)$ of $c$ such that $f(0)=0$,
  then there exists a similar function $g(c)$ such that for $\cN$,
  \[
    \limsup_{n\rightarrow\infty} \frac{1}{n}\log N_E(n,1-2^{-cn}|\cN) \leq Q^{(1)}(\cN) + g(c).
  \]

  In other words, if the error of $\cM$ converges to $1$ exponentially for positive
  rates, then the error of $\cN$ converges to $1$ exponentially for rates
  exceeding $Q^{(1)}(\cN)$.
\end{remark}

\medskip
\begin{remark}
  The type I channel constructed in the proof of Lemma~\ref{lemma:degrading-type-I}
  is such that the composition $UV$ of the Stinespring dilations and of channel
  and degrading channel, actually map the input space $A'$ isometrically into
  $F\ox\Sym^2(E) \subset F\ox E\ox E'$, so that $X_F=\1$.
  
  Looking at Lemma~\ref{lemma:type-I}, we see that the symmetric channel
  constructed there has a dilation $W:G\hookrightarrow \Sym^2(E) \subset E\ox E'$,
  which is a restriction at the input of the ``universal'' symmetric channel
  $\cS:\cL(\Sym^2(E)) \rightarrow \cL(E)$ with the trivial Stinespring dilation
  \[
    \Sym^2(E) \hookrightarrow E \ox E'.
  \]
  To prove a full strong converse for all degradable channels,
  by Theorem~\ref{thm:Q-strong:from-symmetric-to-degradable} it is thus enough
  to show the strong converse for the channels $\cS$, for arbitrarily large
  dimension $|E|$. More precisely, $|E| = 2|A||B|$ is enough for all degradable
  channels with given input and output spaces $A$ and $B$.
\end{remark}

\section{A semidefinite programming approach to the min-entropy of multiply symmetric states}
\label{sec:SDP}
In the proof of Theorem~\ref{thm:Q-semi-strong} we came across a 
term $-H_{\max}^{\epsilon'}(AF^n|{E'}^n)$, $\epsilon'$ being larger than
the coding error we want to analyze. Similarly, in the proof of
Theorem~\ref{thm:P-semi-strong} we had $-H_{\max}^{\epsilon'}(F^n|{E'}^nX)$.

In both cases, assuming w.l.o.g.~that the channel $\cN$ is of type I
(Lemma~\ref{lemma:degrading-type-I}) and using Lemma~\ref{lemma:type-I},
we may view both expressions as 
$-H_{\max}^{\epsilon'}(G^n|{E'}^n) = H_{\min}^{\epsilon'}(G^n|E^n)$,
with respect to an input-output joint state of a symmetric channel 
$\cM^{\ox n}$. Lemma~\ref{lemma:type-I} also informs us that $\cM$ (or
a trivial modification of $\cM$) has a Stinespring dilation 
$W:G \hookrightarrow \Sym^2(E) \subset E\ox E'$; 
in fact, w.l.o.g.~$G=\Sym^2(E)$ but we will not use this.

Now, in the proofs of Theorems~\ref{thm:Q-semi-strong} and \ref{thm:P-semi-strong}
we only made use of the fact that $\cM^{\ox n}$ is symmetric with respect
to exchanging the entire output with the entire environment system.
This symmetry was enough to show that for $\epsilon' < \threshold$ 
this term can bounded by a constant; we also remarked that for larger
$\epsilon'$ this kind of argument cannot be applied.

However, it is obvious that the channel has much more structure, which we
ought to exploit.
Indeed, it is symmetric with respect to exchanging the output and environment
systems of any subset of the $n$ instances of $\cM$ while leaving the
others in place, i.e.~for any $I\subset [n]$,
\[
  \SWAP_{EE'}^{\ox I} W^{\ox n} = W^{\ox n},
\]
and so the joint state of input, output and environment,
$\ket{\psi}^{G^nE^n{E'}^n} = (\1\ox W)^{\ox n}\ket{\phi}^{G^n{G'}^n}$,
satisfies similarly
\begin{equation}\begin{split}
  \label{eq:Z2-n-symmetry}
  \SWAP_{EE'}^{\ox I}{\psi}^{G^nE^n{E'}^n} 
      &= {\psi}^{G^nE^n{E'}^n}                    \\
      &= {\psi}^{G^nE^n{E'}^n}\SWAP_{EE'}^{\ox I},
\end{split}\end{equation}
for all subsets $I$.

The semidefinite programming (SDP) formulation for the smoothed min-entropy
is given by (cf.~\cite{VDTR12})
\[\begin{split}
  2^{-H_{\min}^{\epsilon'}(G^n|E^n)} &= \min\ \tr \sigma^{E^n} \ \text{ s.t.} \\
            &\phantom{= \min}
             \rho^{G^nE^n{E'}^n} \geq 0,\ \tr \rho \leq 1, \\
            &\phantom{= \min}
             \tr \rho\psi \geq 1-{\epsilon'}^2 =: \delta, \\
            &\phantom{= \min}
             \rho^{G^nE^n} \leq \1^{G^n} \ox \sigma^{E^n}.
\end{split}\]
By duality theory (cf.~\cite{VDTR12}) this value is equal to the dual SDP,
given by
\[\begin{split}
  2^{-H_{\min}^{\epsilon'}(G^n|E^n)} &= \max\ \delta r - s \ \text{ s.t.} \\
            &\phantom{= \max}
             r,\,s \geq 0,\ X^{G^nE^n} \geq 0, \\
            &\phantom{= \max}
             r\psi^{G^nE^n{E'}^n} \leq X^{G^nE^n} \ox \1^{{E'}^n} + s\1, \\
            &\phantom{= \max}
             \tr_{G^n} X \leq \1^{E^n}.
\end{split}\]
Note that we get an upper bound on $H_{\min}^{\epsilon'}(G^n|E^n)$
from every dual feasible point (a triple $r,s,X$). The problem is
to construct such a dual feasible point for each pure state ${\psi}^{G^nE^n{E'}^n}$
with the symmetries (\ref{eq:Z2-n-symmetry}) and each $\delta > 0$, such that 
$\delta r - s \geq 2^{-\Omega(\sqrt{n})}$.
Since so far we were unable to find such a construction, we leave the
problem at this point to the attention of the reader.

\section{Conclusion}
\label{sec:conclusion}
For degradable quantum channels, whose quantum and private capacities are
known to be given by the single-letter maximization of the coherent information (which is
then also additive on the class of all degradable channels), we have shown
how to use the powerful min- and max-entropy calculus to derive bounds on the 
optimal quantum and private classical rate, for every finite blocklength $n$.
These bounds improve on the well-known weak converse in that they give
asymptotically the capacity as soon as the error (parametrized by the purified
distance) is small enough: for $Q$ this was $\threshold$, the error of
a 50\%-50\% erasure channel, for $P$ we could get $\frac{1}{3\sqrt{2}}$.
Since this says equivalently that the minimum attainable error jumps from
$0$ to at least some threshold as the coding rate increases above the capacity,
we speak of a ``\semistrong{}'' converse (halfway between a weak and a proper
strong converse).

We have shown furthermore that it is enough to prove a strong converse for
certain universal symmetric (degradable and anti-degradable) channels,
namely those whose Stinespring dilation is the embedding of $\Sym^2(E)$
into $E\ox E'$ as a subspace; then the strong converse would follow for
all degradable channels.
To deal with these symmetric channels, and more
generally with states exhibiting $n$-fold exchange symmetry between
output and environment systems, we discussed briefly a semidefinite
programming (SDP) approach. The viability of this approach stems from the
fact that bounding the relevant min-entropy can be cast as a dual
SDP, and so upper bounds may be obtained by any single dual feasible point.
We have not been able to carry this part of the programme through yet.

Note that the proofs use the quantum AEP, but this does not mean that these
results are restricted to i.i.d.~channels. In fact, by using a standard
discretization argument one can prove that for an arbitrary non-stationary
memoryless channel $\cN_1 \ox \cdots \ox \cN_n$, where each 
$\cN_t:\cL(A)\rightarrow\cL(B)$ is degradable, and sufficiently small error,
the obviously defined $\log N(n,\epsilon)$, $\log N_E(n,\epsilon)$
and $\log M(n,\epsilon,\delta)$ are asymptotically
$\sum_{t=1}^n Q^{(1)}(\cN_t) \pm o(n)$ --- cf.~\cite{Ahlswede:paper-0}
and \cite{Winter:PhD} for analogous statements for 
classical and classical-quantum channels, respectively.

Most channels of course are not degradable (or anti-degradable). 
For practically all these others 
we do not have any approach to obtain a strong or even just a \semistrong{}
converse. One might speculate that other channels with additive
coherent information, hence with a single-letter capacity formula, are also
amenable to our method. But already the very attractive-looking
class of \emph{conjugate degradable} channels~\cite{conjugate-degradable} 
poses new difficulties. 

A related but different question is whether the 
\emph{symmetric side channel-assisted quantum capacity} $Q_{\rm ss}(\cN)$~\cite{SSW},
which has an additive single-letter formula, obeys a \semistrong{} converse.
Note that since arbitrary symmetric side-channels are permitted, including
arbitrarily large 50\%-50\% erasure channels, the strong converse
cannot hold for this capacity, since even infinite rate is achievable with
error $\threshold$. Our present techniques, requiring bounds on the various
system dimensions of the channel, do not to apply, and we seem to need new ideas.

\medskip
{\bf Note on related work.} 
In \cite{SharmaWarsi}, Sharma and Warsi show that one may formulate upper bounds
on the fidelity of codes in terms of the rate and so-called generalized
divergences. Their approach doesn't appear to be related to ours, but it
is conceivable that it may lead to proofs of strong converses for certain
channels' quantum capacity. This however seems to presuppose that channel
parameters derived from these divergences have strong additivity
properties, which can only hold for channels with additive coherent 
information. 

More precisely, the upper bound on the fidelity contained in~\cite[Thm.~1]{SharmaWarsi}
is of no direct use, much as the trivial first steps in the proofs 
of our Theorems~\ref{thm:Q-semi-strong} and~\ref{thm:P-semi-strong}. The
reason is that the bound explicitly depends on the code, via the joint 
input-output state. The only hope at this point is to control the maximum
of said bound over all such input-output states.
It is natural to expect that an important step might be to show that the
maximum is attained on product states.
Crucially, the nature of the maximum bound is not addressed in~\cite{SharmaWarsi}.
Instead it is shown for the quantum erasure channel, that the bound, evaluated 
on the input-output state corresponding to maximally mixed input (which is 
indeed a tensor power), decreases exponentially.

This is the meaning of~\cite[Thm.~3]{SharmaWarsi}, as one can discover from
the calculation following its statement. Literally however, it says
``The strong converse holds for the quantum erasure
  channel for the maximally entangled channel inputs'', 
which might lead an unsuspecting reader to believe that indeed the strong converse
is proved there, albeit perhaps with some restriction that is left vague.
The concluding paragraph unfortunately repeats this claim in the 
stronger words
``To summarize our results, we have given an exponential upper bound on the
  reliability of quantum information transmission'',
and
``We then apply our bound to yield the first known example for exponential 
  decay of reliability at rates above the capacity for quantum information
  transmission''. 
Nothing could be further from the truth; not a single instance of exponential
decay of fidelity above the capacity has been shown within the approach 
of~\cite{SharmaWarsi}. This is because the dependence on $n$ of the maximum bound
in~\cite[Thm.~1]{SharmaWarsi} is not generally understood for any code
family large enough to include capacity achieving codes.

Indeed, claims such as the ones quoted above, would necessarily have to 
involve a bound on all conceivable quantum codes, for large $n$, which 
seems difficult, to say the least. 
But the only code that~\cite[Thm.~3]{SharmaWarsi} covers is the trivial 
one of using the entire input bandwidth, not encoding at all. 
To analyze it, however, one 
hardly needs the machinery developed in~\cite{SharmaWarsi}; the
reader may wish to convince her-/himself that \emph{every} noisy
channel exhibits exponential decay of fidelity for this code.

\section*{Acknowledgments}
We thank Mario Berta and Marco Tomamichel for discussions on strong converses
in the context of quantum data compression with side information, Robert K\"onig and
Stephanie Wehner for illuminating comments on strong converses, and Renato
Renner and Fr\'ed\'eric Dupuis for sharing with us many of their 
insights regarding min-, max- and other entropies. In particular, we gratefully
acknowledge Fr\'ed\'eric Dupuis' permission to use his result on the comparison
between smooth max- and min-entropy (Lemma~\ref{lemma:max-min-inequality});
and Robert K\"onig's suggestion of the name ``\semistrong{} converse'',
as well as the PPT example in Section~\ref{sec:converses}.
Normand Beaudry and Mark Wilde, as well as the anonymous referees,
kindly suggested several improvements over a the original preprint version.

\appendix

\section{Proofs of lemmas and propositions}
\label{app:proofs}
Here we present the proofs of several auxiliary results used in the proof
of the main result, which would have broken the flow of the text.

\medskip
\begin{IEEEproof}[Proof of Lemma~\ref{lemma:concavity}]
Define the auxiliary state
\[
  \overline{\rho}^{ABX} := \sum_i p_i \rho_i^{AB} \ox \proj{i}^X,
\]
so that the average of the $\rho_i$ becomes
$\overline{\rho}^{AB} = \tr_X \overline{\rho}^{ABX}$. Choosing
purifications $\psi_i^{ABC}$, we can consider the following purification
of $\overline{\rho}^{ABX}$:
\[
  \ket{\varphi}^{ABCXY} = \sum_i \sqrt{p_i} \ket{\psi_i}^{ABC} \ket{i}^X \ket{i}^Y.
\]
Then, using monotonicity (Lemma~\ref{lemma:mono}) and duality,
\begin{equation}\begin{split}
  \label{eq:step-1}
  H_{\max}^\epsilon(A|B)_{\overline{\rho}^{AB}}
      &\geq H_{\max}^\epsilon(A|BX)_{\overline{\rho}^{ABX}} \\
      &=    -H_{\min}^\epsilon(A|CY)_{\varphi},
\end{split}\end{equation}
observing $\varphi^{ACY} = \sum_i p_i \psi_i^{AC} \ox \proj{i}^Y$.

Now, by definition of the smooth min-entropy, its exponential is give by
the following optimization:
\[\begin{split}
  \Phi_\epsilon(\varphi^{A:CY}) &:= 2^{-H_{\min}^\epsilon(A|CY)_{\varphi}} \\
                                &=  \min \tr \sigma^{CY} \text{ s.t. } \\
                                &\phantom{= \min}
                                 \widetilde{\rho}^{ACY} \leq \1^A \ox \sigma^{CY}, \\
                                &\phantom{= \min}
                                 \widetilde{\rho} \geq 0,\ \tr\widetilde{\rho} \leq 1, \\
                                &\phantom{= \min}
                                 F(\varphi^{ACY},\widetilde{\rho}) 
                                     =    \|\sqrt{\varphi}\sqrt{\widetilde{\rho}}\|_1 
                                     \geq \sqrt{1-\epsilon^2}.
\end{split}\]
Since $\varphi^{ACY}$ is invariant under phase unitaries on $Y$, we may assume
w.l.o.g.~that both $\widetilde{\rho}$ and $\sigma$ have the same property,
i.e.~they may be assumed to be classical on $Y$:
\begin{align*}
  \widetilde{\rho}^{ACY} &= \sum_i q_i \widetilde{\rho}_i^{AC} \ox \proj{i}^Y, \\
  \sigma^{CY}            &= \sum_i q_i \sigma_i^C \ox \proj{i},
\end{align*}
where $q_i \geq 0$, $\sum_i q_i = 1$ and $\widetilde{\rho}_i \in \cS_{\leq}(AC)$;
furthermore $\sigma_i \geq 0$. With these notations, the objective function
in the above optimization is $\tr \sigma^{CY} = \sum_i q_i \tr\sigma_i^C$,
the first constraint is equivalent to $\widetilde{\rho}_i^{AC} \leq \1^A\ox\sigma_i^C$
for all $i$, and 
\[
  F(\varphi^{ACY},\widetilde{\rho}) 
    = \sum_i \sqrt{p_i q_i} F(\psi_i^{AC},\widetilde{\rho}_i^{AC}).
\]
Thus, observing that the $\psi_i^{AC}$ are related to $\psi_1^{AC} = \tr_B \psi_1$
by local unitaries, we have
\[\begin{split}
  \Phi_\epsilon(\varphi^{ACY}) &= \min \sum_i q_i \Phi_{\epsilon_i}(\psi_i^{AC}) \text{ s.t.}\\
                               &\phantom{= \min}
                                \sum_i \sqrt{p_i q_i}\sqrt{1-\epsilon_i^2} \geq \sqrt{1-\epsilon^2} \\
                               &= \min \sum_i q_i \Phi_{\epsilon_i}(\psi_1^{AC}) \text{ s.t.}\\
                               &\phantom{= \min}
                                \sum_i \sqrt{p_i q_i}\sqrt{1-\epsilon_i^2} \geq \sqrt{1-\epsilon^2},
\end{split}\]
where the variables are $q_i$ and $\epsilon_i$.

Now, Cauchy-Schwarz inequality says
\[
  \sum_i \sqrt{p_i q_i}\sqrt{1-\epsilon_i^2}
     \leq \sqrt{\sum_i p_i\sqrt{1-\epsilon_i^2}} \sqrt{\sum_i q_i\sqrt{1-\epsilon_i^2}}.
\]
Hence the constraint implies that $\sum_i q_i\sqrt{1-\epsilon_i^2} \geq 1-\epsilon^2$
and we get
\[\begin{split}
  \Phi_\epsilon(\varphi^{ACY}) &\geq \min \sum_i q_i \Phi_{\epsilon_i}(\psi_1^{AC}) \text{ s.t.}\\
                               &\phantom{\geq \min}
                                \sum_i q_i\sqrt{1-\epsilon_i^2} \geq 1-\epsilon^2. \\
\end{split}\]
For each $i$, $\Phi_{\epsilon_i}(\psi_1^{AC}) = \tr \sigma_i^C$ with
$0 \leq \widetilde{\rho}_i \leq \1^A \ox \sigma_i^C$, $\tr\widetilde{\rho}_i \leq 1$,
and $F(\psi_1^{AC},\widetilde{\rho}_i) \geq \sqrt{1-\epsilon_i^2}$.
Thus, forming $\widetilde{\omega} := \sum_i q_i \widetilde{\rho}_i \in \cS(AC)$ 
and $\widetilde{\sigma} = \sum_i q_i \sigma \geq 0$, we have $\tr\widetilde{\omega}\leq 1$,
$\widetilde{\omega} \leq \1 \ox \widetilde{\sigma}$ and
\[
  F(\psi_1^{AC},\widetilde{\omega}) \geq \sum_i q_i \sqrt{1-\epsilon_i^2}
                                    \geq 1-\epsilon^2 
                                    =: \sqrt{1-\widehat{\epsilon}^2},
\]
where $\widehat{\epsilon} \leq \epsilon\sqrt{2}$.

This gives eventually
\[
  \Phi_{\epsilon}(\varphi^{A:CY}) \geq \Phi_{\widehat{\epsilon}}(\psi_1^{AC}),
\]
so going back to Eq.~(\ref{eq:step-1}), we arrive at
\[\begin{split}
  H_{\max}^\epsilon(A|B)_{\overline{\rho}} &\geq -H_{\min}^\epsilon(A|CY)_\varphi             \\
                                           &\geq -H_{\min}^{\widehat{\epsilon}}(A|C)_{\psi_1} \\
                                           &=     H_{\max}^{\widehat{\epsilon}}(A|C)_\rho     \\
                                           &\geq  H_{\max}^{\epsilon\sqrt{2}}(A|C)_\rho,
\end{split}\]
and we are done.
\end{IEEEproof}

\medskip
\begin{IEEEproof}[Proof of Lemma~\ref{lemma:max+log-bound}]
Fix purifications $\psi_i^{ABC}$ of the $\rho_i$, so that $\overline{\rho}$
can be purified as
\[
  \ket{\psi}^{ABCC_0} = \sum_{i=1}^M \sqrt{p_i} \ket{\psi_i}^{ABC} \ket{i}^{C_0}.
\]
We use the following characterization of smooth max-entropies
(cf.~\cite{TomamichelThesis}):
\[\begin{split}
  2^{H_{\max}^\epsilon(A|B)_{\rho_i}} &= \min \| \tr_A Z_i \| \text{ s.t.} \\
                                      &\phantom{= \min}
                                       F(\psi_i,\psi_i') \geq \sqrt{1-\epsilon^2}, \\
                                      &\phantom{= \min}
                                       \psi_i' \leq Z_i^{AB} \ox \1^C.
\end{split}\]

Fix optimal $\ket{\psi_i'} \in ABC$, such that
$\bra{\psi_i}\psi_i'\rangle = F(\psi_i,\psi_i') \geq \sqrt{1-\epsilon^2}$,
and $Z_i \geq 0$. Let $\lambda = \max_i \| \tr_A Z_i \|$ and define
\[
  \ket{\psi'}^{ABCC_0} := \sum_{i=1}^M \sqrt{p_i} \ket{\psi_i'}^{ABC} \ket{i}^{C_0},
\]
so that 
\[
  F(\psi,\psi') = \bra{\psi}\psi'\rangle 
                = \sum_i p_i \bra{\psi_i}\psi_i'\rangle
                \geq \sqrt{1-\epsilon^2}.
\]
Furthermore, using Hayashi's pinching inequality~\cite{Hayashi,OgawaHayashi} 
in the second line,
\[\begin{split}
 \proj{\psi'} &=      \sum_{ij=1}^M \sqrt{p_i p_j} \ket{\psi_i'}\!\bra{\psi_j'} \ox \ket{i}\!\bra{j} \\
              &\leq M \sum_{i=1}^M p_i {\psi_i'}^{ABC} \ox \proj{i}^{C_0} \\
              &\leq \sum_i M p_i Z_i^{AB} \ox \1^C \ox \1^{C_0} \\
              &=:   Z^{AB} \ox \1^{CC_0}.
\end{split}\]

I.e., $\psi'$ and $Z$ are feasible for $\overline{\rho}$, and 
the objective function value
\[\begin{split}
  \| \tr_A Z \| &=    \left\| \sum_i M p_i \tr_A Z_i \right\| \\
                &\leq \sum_i M p_i \| \tr_A Z_i \| \\
                &\leq M\lambda
\end{split}\]
gives an upper bound to $2^{H_{\max}^\epsilon(A|B)_{\overline{\rho}}}$.
Thus we can conclude
\[\begin{split}
  H_{\max}^\epsilon(A|B)_{\overline{\rho}} &\leq \log\lambda + \log M \\
                                           &=    \max_i H_{\max}^{\epsilon}(A|B)_{\rho_i} + \log M,
\end{split}\]
as advertised.
\end{IEEEproof}

\medskip
\begin{IEEEproof}[Proof of Proposition~\ref{prop:AEP}]
To get bounds valid for all $n$, we use well-known tail estimates for
sums of independent random variables due to Hoeffding~\cite{DemboZeitouni}.
Namely, consider the discrete random variable $X$ with minimum non-zero
probability $\min_x P_X(x) =: 2^{-\mu}$ and let $L = L(X) := -\log P_X(X)$,
such that $0\leq L\leq \mu$ with probability $1$, and $\EE L = H(P)$.
Then, for i.i.d.~realizations $X_1,X_2,\ldots,X_n$ of $X$, and associated
$L_i$, Hoeffding's inequality states
\begin{equation}\begin{split}
  \Pr\left\{ \sum_{i=1}^n L_i > n H(P) + \Delta\sqrt{n} \right\} &\leq e^{-\frac{2\Delta^2}{\mu^2}}, \\
  \Pr\left\{ \sum_{i=1}^n L_i < n H(P) - \Delta\sqrt{n} \right\} &\leq e^{-\frac{2\Delta^2}{\mu^2}}.
  \label{eq:The-Hoff}
\end{split}\end{equation}

We can use these bounds to construct typical projectors for a state $\rho^{\ox n}$,
$\rho \in \cS(\cH)$, in the usual way.
Let $\rho = \sum_x \lambda_x \proj{x}$ be a diagonalization,
so that $\lambda_x$ can be interpreted as a probability distribution on the $x$.
Define two projectors
\begin{align*}
  P_{\rho^{\ox n}}^{+\Delta} &:= \sum_{x^n \in \cT_{\lambda^{\ox n}}^{+\Delta}} \proj{x^n} \text{ with} \\
  \cT_{\lambda^{\ox n}}^{+\Delta} &:= \left\{ x^n=x_1\ldots x_n : 
                                       \sum_i -\log\lambda_{x_i} \leq n S(\rho) + \Delta\sqrt{n} \right\},
\end{align*}
and
\begin{align*}
  P_{\rho^{\ox n}}^{-\Delta} &:= \sum_{x^n \in \cT_{\lambda^{\ox n}}^{-\Delta}} \proj{x^n} \text{ with} \\ 
  \cT_{\lambda^{\ox n}}^{-\Delta} &:= \left\{ x^n=x_1\ldots x_n : 
                                       \sum_i -\log\lambda_{x_i} \geq n S(\rho) - \Delta\sqrt{n} \right\}.\end{align*}
By Eq.~(\ref{eq:The-Hoff}),
\begin{align*}
  \tr \rho^{\ox n}P_{\rho^{\ox n}}^{+\Delta} &\geq 1-e^{-\frac{2\Delta^2}{\mu^2}}, \\
  \tr \rho^{\ox n}P_{\rho^{\ox n}}^{-\Delta} &\geq 1-e^{-\frac{2\Delta^2}{\mu^2}},
\end{align*}
where $\mu = \log\|\rho^{-1}\|$.

Now, for a pure tripartite state $\ket{\psi}\in ABC$, let $\Delta > 0$ and consider
the projectors
\begin{align*}
  P_B^+ &:= P_{\rho_B^{\ox n}}^{+\Delta\mu_B}, \\
  P_C^- &:= P_{\rho_C^{\ox n}}^{-\Delta\mu_C}.
\end{align*}
Defining $\ket{\Psi'} := (\1_A \ox P_B^+ \ox P_C^-)\ket{\psi}^{\ox n}$, clearly we have
\[\begin{split}
  \bra{\Psi'}\psi\rangle^{\ox n} &=    \bra{\psi}^{\ox n} (\1_A \ox P_B^+ \ox P_C^-) \ket{\psi}^{\ox n} \\
                                 &\geq 1-2e^{-2\Delta^2}                                \\
                                 &\stackrel{!}{\geq} \sqrt{1-\epsilon^2},
\end{split}\]
for $\Delta = \sqrt{\ln\frac{2}{\epsilon}}$. By definition
\[
  {\Psi'}^{C^n} \simeq {\Psi'}^{A^nB^n} \leq 2^{-n S(AB) + \Delta\mu_C\sqrt{n}}(\1^{A^n}\ox P_B^+).
\]
On the other hand, we just need to rescale $P_B^+$ by its trace,
$\sigma := \frac{1}{\tr P_B^+} P_B^+$ to get an eligible state in the definition
of $H_{\min}^\epsilon(A|B)$. Note that $\tr P_B^+ \leq 2^{n S(\rho) + \Delta\mu_B\sqrt{n}}$,
hence
\[
  {\Psi'}^{A^nB^n} \leq 2^{-n S(A|B) + \Delta(\mu_B+\mu_C)\sqrt{n}} (\1^{A^n}\ox\sigma^{B^n}),
\]
thus showing
\[
  H_{\min}^\epsilon(A|B) \geq n S(A|B) - (\mu_B+\mu_C)\sqrt{n\ln\frac{2}{\epsilon}}.
\]
The upper bound on $H_{\max}^\epsilon(A|B)$ follows by the duality
of the min- and max-entropies, as well as that of the conditional 
von Neumann entropy: $S(A|B) = -S(A|C)$.
\end{IEEEproof}

\medskip
\begin{IEEEproof}[Proof of Proposition~\ref{prop:one-shot-quantum-capacity}]
For a $d$-dimensional projector $Q$ on $A$, write 
\[
  \sqrt{\frac{|A|}{d}}(Q\ox\1)\ket{\psi^{ABE}} =: \sqrt{t_Q} \ket{\widetilde{\psi}_Q}^{ABE},
\]
where $\sqrt{t_Q}$ is the normalisation of the left hand side and $\ket{\widetilde{\psi}_Q}^{ABE}$
is a state. Our goal is to show that we can find $Q$ such that $\widetilde{\psi}_Q^{AE}$ is
close to a product state. To be precise, the claim is that there exists
$\varphi\in\cS_{\leq}(E)$ and $Q$ such that
\begin{equation}
  \label{eq:decoupling}
  P(\widetilde{\psi}_Q,\tau_Q\ox\varphi^E) 
     \leq \eta + 2^{-\frac14\bigl(H_{\min}^\eta(A|E)_\psi-\log d\bigr)}.
\end{equation}
Then, using the familiar decoupling argument, there is a cptp map
$\cD$ acting on $B$ such that
\[
  P\bigl( (\id\ox\cD)\widetilde{\psi}_Q^AB, \Phi_{QQ'} \bigr) 
     \leq \eta + 2^{-\frac14\bigl(H_{\min}^\eta(A|E)_\psi-\log d\bigr)},
\]
where $\Phi_{QQ'}$ is a maximally entangled state. Choosing 
\[
  \ket{\widetilde{\phi}_Q}^{AA'} := \sqrt{|A|}{d t_Q}(Q\ox \1)\ket{\phi}
\]
as the input state, so that
$\ket{\widetilde{\psi}_Q}^{ABE} = (\1\ox U)\ket{\widetilde{\phi}_Q}$,
completes the entanglement-generating code. Choosing
$\log d \leq H_{\min}^\eta(A|E)_\psi - 4\log\frac{1}{\epsilon}$
guarantees that its error is $\leq \eta+\epsilon$.

To prove Eq.~(\ref{eq:decoupling}), choose a $\varphi \in \cS_{\leq}(ABE)$ with
$P(\varphi,\psi)\leq \eta$ and 
$H_{\min}^\eta(A|E)_\psi = H_{\min}(A|E)_\varphi$.
Consider the cptp map
\[
  \cP: \rho \longmapsto \int {\rm d}Q \frac{|A|}{d} Q\rho Q^\dagger \ox \proj{Q},
\]
where $\ket{Q}$ are orthogonal labels of a dummy system.
By the contractiveness of the purified distance, we have
\begin{equation}
  \label{eq:close}
  P\bigl( (\cP\ox\id)\varphi^{AE}, (\cP\ox\id)\psi^{AE} \bigr) \leq \eta.
\end{equation}
We also have $\int {\rm d}Q\ t_Q = 1$.

Now, Lemma~\ref{lemma:Berta} below tells us
\[
  \bigl\| (\cP\ox\id)\varphi^{AE} - (\cP\ox\id)(\tau_A\ox\varphi^E) \bigr\|_1
                                   \leq 2^{-\frac12(H_{\min}(A|E)_\varphi-\log d)},
\]
noting 
\[
  (\cP\ox\id)(\tau_A\ox\varphi^E) = \int {\rm d}Q \tau_Q \ox \varphi^E \ox \proj{Q},
\]
and that the trace norm on the left hand side is
\[
  \int {\rm Q} \left\| \frac{|A|}{d}(Q\ox\1)\varphi^{AE}(Q\ox\1)^\dagger - \tau_Q \ox \varphi^E \right\|_1.
\]
By Eq.~(\ref{eq:P-vs-D}), the trace norm bound implies
\[
   P\bigl( (\cP\ox\id)\varphi^{AE}, (\cP\ox\id)(\tau_A\ox\varphi^E) \bigr)
                                   \leq 2^{-\frac14(H_{\min}(A|E)_\varphi-\log d)}.
\]

Substituting $H_{\min}(A|E)_\varphi = H_{\min}^\eta(A|E)_\psi$ and using 
Eq.~(\ref{eq:close}) with the triangle inequality for the purified distance,
we get
\[\begin{split}
  P&\bigl( (\cP\ox\id)\psi^{AE}, (\cP\ox\id)(\tau_A\ox\varphi^E) \bigr) \\
   &\phantom{==========}  \leq \eta + 2^{-\frac14(H_{\min}^\eta(A|E)_\psi-\log d)} =: \delta.
\end{split}\]
Equivalently, inserting the definition of $\widetilde{\psi}_Q$ and $t_Q$:
\[\begin{split}
  \sqrt{1-\delta^2} &\leq F\bigl( (\cP\ox\id)\psi^{AE}, (\cP\ox\id)(\tau_A\ox\varphi^E) \bigr) \\
          &=    \int {\rm d}Q \left\| \sqrt{t_Q\widetilde{\psi}_Q^{AE}}\sqrt{\tau_Q\ox\varphi^E} \right\|_1 \\
          &=    \int {\rm d}Q \sqrt{t_Q} F(\widetilde{\psi}_Q^{AE},\tau_Q\ox\varphi^E).
\end{split}\]
Since finally, by the concavity of the square root,
\[
  \int {\rm d}Q \sqrt{t_Q} \leq \sqrt{\int {\rm d}Q\ t_Q} = 1,
\]
this implies that there exists $Q$ in the previous integral with
$F(\widetilde{\psi}_Q^{AE},\tau_Q\ox\varphi^E) \geq \sqrt{1-\delta^2}$,
which is precisely Eq.~(\ref{eq:decoupling}).
\end{IEEEproof}

\medskip
\begin{lemma}[Berta~\cite{Berta:dipl}]
\label{lemma:Berta}
Let $\ket{\varphi} \in ABC$ be a state vector. 
Picking a $d$-dimensional projector $Q$ uniformly (i.e. from the
unitarily invariant measure ${\rm d}Q$), we have
\[\begin{split}
  \int {\rm d}Q&\ \left\| \frac{|A|}{d}(Q\ox\1)\varphi^{AE}(Q\ox\1)^\dagger - \tau_Q \ox \psi^E \right\|_1 \\
               &\phantom{============}
                \leq 2^{-\frac12(H_{\min}(A|E)-\log d)},
\end{split}\]
with the maximally mixed state $\tau_Q=\frac{1}{d}Q \in \cS(A)$ on the support of $Q$.
\altqed
\end{lemma}

\end{document}